\documentclass[Journal,a4paper]{IEEEtran}

\usepackage[T1]{fontenc}% optional T1 font encoding
\usepackage{graphicx}
\usepackage{setspace}
\usepackage{titlesec}

\titlespacing\section{0pt}{8pt plus 4pt minus 2pt}{4pt plus 2pt minus 2pt}
\titlespacing\subsection{0pt}{6pt plus 4pt minus 2pt}{4pt plus 2pt minus 2pt}
\titlespacing\subsubsection{1em}{6pt plus 4pt minus 2pt}{4pt plus 2pt minus 2pt}
\titleformat{\subsubsection}[runin]
 {\normalfont\normalsize\itshape}
 {\arabic{subsubsection})} {0.5em}{}[:\ ]

\usepackage{amsthm,amsmath,amssymb,bm}
\usepackage{mathrsfs}
\usepackage{amsfonts}

\newtheorem{lemma}{Lemma}

\newtheorem{proposition}{Proposition}
\usepackage[caption=false]{subfig}
\usepackage{algorithm}
\usepackage[noend]{algpseudocode}
\usepackage{graphics}
\usepackage{epsfig}
\usepackage{cite}
\usepackage{color}
\usepackage{booktabs}
\usepackage{verbatim}
\usepackage{stfloats}
\usepackage[colorlinks,linkcolor=blue,urlcolor=blue,citecolor=blue]{hyperref}
\usepackage{cases}
\usepackage{multirow}
\usepackage{enumitem}
\usepackage{ragged2e}
\makeatletter
\renewcommand{\maketag@@@}[1]{\hbox{\m@th\normalsize\normalfont#1}}%
\makeatother

\makeatletter
\renewcommand{\theALG@line}{\arabic{ALG@line}}
\newcounter{subline}[ALG@line]

\newlength{\algnumwidth}
\newcommand{\subState}[1]{%
    \refstepcounter{subline}%
    \Statex
    \noindent
    \begin{tabular}{@{}p{\algnumwidth}@{}}
        \raggedleft\footnotesize\theALG@line.\arabic{subline}:
    \end{tabular}%
    \hspace{-\algnumwidth}%
    \hspace{\algorithmicindent}%
 #1%
}

\makeatother

\newcounter{MYtempeqncnt}

\newcommand{\transpose}{\mathsf{T}}
\newcommand{\hermconj}{\mathsf{H}}
\newcommand{\trace}{\mathtt{tr}}
\newcommand{\vect}{\mathrm{vec}}
\begin{document}
    
\title{Joint Beamforming and Antenna Position Optimization for Fluid Antenna-Assisted MU-MIMO Networks}
\author{Tianyi~Liao,~\IEEEmembership{Graduate~Student Member,~IEEE},~Wei~Guo,~\IEEEmembership{Member,~IEEE},~Hengtao~He,~\IEEEmembership{Member,~IEEE},\\ Shenghui~Song,~\IEEEmembership{Senior Member,~IEEE},~Jun~Zhang,~\IEEEmembership{Fellow,~IEEE},~Khaled B. Letaief,~\IEEEmembership{Fellow,~IEEE}\vspace{-1cm}
\thanks{
    This work was supported by the Hong Kong Research Grants Council (RGC) under the AoE Grant AoE/E-601/22R, in part by the General Research Fund (Project No. 16209524) from the Hong Kong RGC, and in part by the National Natural Science Foundation of China (NSFC) under Grants 62501144. An earlier version of this paper was presented in part at the 2025 IEEE Global Communication Conference, Taipei, Taiwan, Dec. 2025~\cite{liao2025fluid}. \textit{(Corresponding author: Wei Guo.)}

    Tianyi Liao, Wei Guo, Jun Zhang, and Khaled B. Letaief are with the Department of Electronic and Computer Engineering, The Hong Kong University of Science and Technology (HKUST), Hong Kong. (Emails: ty.liao@connect.ust.hk, \{eeweiguo, eejzhang, eekhaled\}@ust.hk).

    Hengtao He is with the School of Information Science and Engineering, Southeast University, Nanjing, China. (Email: hehengtao@seu.edu.cn).

    Shenghui Song is with the Division of Integrative Systems and Design and the Department of Electronic and Computer Engineering, HKUST, Hong Kong. (Email: eeshsong@ust.hk).

	The source code is publicly available at https://github.com/liaotianyi0114/FAS-MIMO-WSR-Maximization.
}}

\maketitle
\IEEEaftertitletext{\vspace{-1\baselineskip}}

\begin{abstract}
    The fluid antenna system (FAS) is a disruptive technology for future wireless communication networks. This paper considers the joint optimization of beamforming matrices and antenna positions for weighted sum rate (WSR) maximization in fluid antenna (FA)-assisted multiuser multiple-input multiple-output (MU-MIMO) networks, which presents significant challenges due to the strong coupling between beamforming and FA positions, the non-concavity of the WSR objective function, and high computational complexity. To address these challenges, we first propose a novel block coordinate ascent (BCA)-based method that employs matrix fractional programming techniques to reformulate the original complex problem into a more tractable form. Then, we develop a \emph{parallel} majorization maximization (MM) algorithm capable of optimizing all FA positions simultaneously. To further reduce computational costs, we propose a decentralized implementation based on the decentralized baseband processing (DBP) architecture. Simulation results demonstrate that our proposed algorithm not only achieves significant WSR improvements over conventional MIMO networks but also outperforms the existing method. Moreover, the decentralized implementation substantially reduces computation time while maintaining similar performance compared with the centralized implementation.
\end{abstract} 

\begin{IEEEkeywords}
Fluid antenna system (FAS), MU-MIMO, fractional programming (FP), majorization maximization (MM), decentralized baseband processing (DBP).
\end{IEEEkeywords}

\maketitle

\section{Introduction}\label{sec:intro}
\IEEEPARstart{T}{he} sixth-generation (6G) wireless communication systems aim to achieve terabit-per-second data rates, high energy efficiency, and sub-millisecond latency~\cite{letaiefRoadmap6GAI2019,saadVision6GWireless2020,wang2023road}. To achieve these objectives, massive multiple-input multiple-output (MIMO) and multiuser MIMO (MU-MIMO) technologies~\cite{lu2014overview,foschini1998limits} will play pivotal roles. The major advantage of MIMO systems is their ability to leverage spatial degrees of freedom (DoF), which can improve the system performance by exploiting spatial diversity and multiplexing gains~\cite{zhengDiversityMultiplexingFundamental2003}. However, conventional MIMO systems typically assume fixed-position antenna (FPA) configurations, which lack adaptability to varying propagation environments and thus cannot fully exploit the spatial DoF. Specifically, some antennas inevitably suffer from deep fading and may be subject to strong interference, leading to significant degradation in the signal-to-interference-plus-noise ratio (SINR).

To address these challenges, the fluid antenna system (FAS) was proposed in~\cite{wong2020fluid} as a disruptive technology. FASs leverage fluid antennas (FAs) to enable flexible control over antenna positions, gains, radiation patterns, and other key characteristics~\cite{wongFluidAntennaSystems2021}. Among these capabilities, position reconfiguration has proven to be an effective means of fully exploiting spatial DoF. Existing implementations of position-reconfigurable FASs include pixel-based~\cite{song2013efficient,chai2022port,waqar2023deep}, liquid-based~\cite{martinez2022toward,wang2022continuous}, and motor-driven designs~\cite{ma2023capacity,maMIMOCapacityCharacterization2024,zhuModelingPerformanceAnalysis2024}.

Position-reconfigurable FASs can be broadly categorized into pixel-based~\cite{song2013efficient,chai2022port,waqar2023deep}, liquid-based~\cite{martinez2022toward,wang2022continuous}, and motor-driven implementations~\cite{ma2023capacity,maMIMOCapacityCharacterization2024,zhuModelingPerformanceAnalysis2024}.
Among these categories, motor-driven FASs offer higher reconfiguration fidelity and can be seamlessly integrated into MIMO systems. Considering the movement delay of motor-driven FASs, it is primarily envisioned to be deployed in ultra massive machine-type communication (umMTC) scenarios where the surrounding environment varies slowly~\cite{zhuMovableAntennasWireless2024}.\footnote{Motor-driven FASs also face other implementation issues, including potentially excessive motor power consumption and calibration errors. Nevertheless, recent works have attempted to alleviate movement delay~\cite{wangThroughputMaximizationMovable2025} and to jointly optimize motor power consumption~\cite{weiMechanicalPowerModeling2025}. We do not consider these implementation issues, as the focus of this paper is on analyzing the performance limits of FASs.} The advantages of incorporating FAs into MU-MIMO networks are twofold. First, the positions of FAs can be dynamically adjusted to avoid deep fading in desired links, thereby enhancing the optimal diversity–multiplexing tradeoff~\cite{newInformationTheoreticCharacterizationMIMOFAS2024}. Second, FAs help mitigate multiuser interference (MUI) by optimizing antenna locations, since interference can experience deep fading with appropriate position adjustments~\cite{wongOpportunisticFluidAntenna2023,zhuMovableAntennaEnhancedMultiuser2024,newTutorialFluidAntenna2024}.

\subsection{Motivation}\label{subsec:motivation}
The weighted sum rate (WSR) maximization problem is essential to optimize the overall system capacity in MU-MIMO systems~\cite{christensenWeightedSumrateMaximization2008,bogaleWeightedSumRate2012}, which has been extensively studied in the context of conventional MIMO systems~\cite{shen2018fractional,shen2018fractional2,shenGraphNeuralNetworks2023}. However, the WSR maximization problem involving joint beamforming and FA position optimization in FA-assisted MU-MIMO systems poses significant and unprecedented challenges. First, beamforming and FA positions are strongly coupled. On one hand, beamforming alters the effective path coefficients of the channels, making the optimal FA positions dependent on the beamforming matrices. On the other hand, the array manifold, which is directly affected by the FA positions, plays a critical role in shaping the array beam pattern~\cite{balanis2016antenna}. Although the method in~\cite{fengWeightedSumRateMaximization2024} effectively decouples beamforming and FA positions, it cannot be readily extended to MU-MIMO setups. Second, the relationship between the channel characteristics and FA positions is highly non-linear, rendering the WSR maximization problem non-convex. In particular, small-scale fading induces rapid and irregular variations in the channel with respect to (w.r.t.) the FA positions~\cite{zhuModelingPerformanceAnalysis2024}, posing significant challenges for accurately determining the optimal FA configuration. To cope with such non-convexity, existing works often resort to iterative optimization techniques such as successive concave approximation (SCA)~\cite{maMIMOCapacityCharacterization2024} and majorization–maximization (MM)~\cite{tangSecureMIMOCommunication2025,fengWeightedSumRateMaximization2024}. However, their performance is largely limited by the tightness of the designed surrogate functions. Last but not least, as a consequence of the non-convexity of the optimization problem, the resulting algorithms are computationally intensive, which hinders scalability to massive MIMO scenarios. These significant challenges highlight the need for novel approaches to maximize the WSR in FA-assisted MU-MIMO systems, thereby motivating the joint design of beamforming and antenna positions investigated in this work.

\subsection{Contribution}
In this paper, we propose a novel algorithm for joint beamforming and FA position optimization in FA-assisted downlink MU-MIMO systems within the block coordinate ascent (BCA) framework. In addition, a decentralized implementation is developed to further reduce computational overhead. The main contributions are summarized as follows.
\begin{itemize}
    \item In contrast to~\cite{fengWeightedSumRateMaximization2024}, which focused on MU-multiple-input single-output (MU-MISO) networks, this work considers the more general FA-assisted MU-MIMO networks. We formulate the joint beamforming and antenna position optimization as a WSR maximization problem, where the objective function is non-concave and the optimization variables are highly coupled. To decouple the beamforming matrices and FA positions, we utilize two \emph{matrix} fractional programming (FP) techniques, i.e., the quadratic transform and the Lagrangian dual transform~\cite{shen2018fractional,shen2018fractional2}.\footnote{It has been shown in~\cite{shen2018fractional,shen2018fractional2} that the WMMSE approach used in~\cite{fengWeightedSumRateMaximization2024} is equivalent to the FP techniques adopted in this paper.} These FP techniques enable a BCA-based algorithm to solve the decoupled subproblems efficiently.
    \item Unlike existing FA position optimization algorithms that sequentially update FA positions, we propose a novel \emph{parallel} optimization algorithm based on the MM framework that simultaneously optimizes all FA positions. A tight surrogate function is constructed using matrix chain rules, providing a more accurate approximation of the original objective. The proposed algorithm outperforms the method in~\cite{fengWeightedSumRateMaximization2024} and can be readily extended to a decentralized implementation to further reduce computational complexity.
    \item To further reduce computational cost, we propose a decentralized implementation of the algorithm using the decentralized baseband processing (DBP) architecture~\cite{li2017decentralized}, which partitions the transmit FA array into multiple clusters. The DBP framework decomposes the optimization problem into smaller subproblems, enabling decentralized units (DUs) to solve them in parallel. To facilitate the design of the decentralized FP-based beamforming algorithm, we adopt the non-homogeneous transform and Nesterov's extrapolation~\cite{zhangDiscerningEnhancingWeighted2023,shenAcceleratingQuadraticTransform2024} to avoid matrix inversion. The proposed MM-based algorithm for FA position optimization can be naturally extended to the decentralized implementation, which substantially reduces computational costs while incurring only negligible performance degradation compared to its centralized counterpart.
\end{itemize}

\subsection{Organization and Notation}\label{subsec:org}
The remainder of this paper is organized as follows. Section~\ref{sec:related} presents the related work. In Section~\ref{sec:system}, the channel model of the FA-assisted MU-MIMO system and the formulation of the WSR maximization problem are provided. Section~\ref{sec:bca_sol} reformulates the problem using FP techniques and solves it using BCA and MM. The decentralized implementation of the proposed algorithm is introduced in Section~\ref{sec:dec_bca}. Simulation results are provided in Section~\ref{sec:sim}, and conclusions are drawn in Section~\ref{sec:conclusion}.

In this paper, $a$, $\mathbf{a}$, and $\mathbf{A}$ denote a scalar, a vector, and a matrix, respectively. The imaginary unit is denoted by $\jmath$. For a complex scalar $a$, its amplitude and phase are given by $\lvert a \rvert$ and $\angle a$, respectively. The $\ell_2$ norm of a vector $\mathbf{a}$ is $\lVert \mathbf{a} \rVert_2$. $[\mathbf{A}]_m$, $[\mathbf{A}]_{mn}$, $\mathbf{A}^\transpose$, $\mathbf{A}^\hermconj$, $\det(\mathbf{A})$, $\trace(\mathbf{A})$, $\vect(\mathbf{A})$, and $\lVert \mathbf{A} \rVert_\infty$ denote the $m$-th row, the $(m, n)$-th element, transpose, conjugate transpose, determinant, trace, vectorization, and the infinity norm of matrix $\mathbf{A}$, respectively. $\mathbf{A} \succeq \mathbf{0}$ and $\mathbf{A} \succ \mathbf{0}$ indicate that $\mathbf{A}$ is positive semi-definite and positive definite, respectively. $\mathbb{C}^{M \times N}$, $\mathbb{R}^{M \times N}$, and $\mathbb{R}_{+}^{M \times N}$ denote the sets of $M \times N$ complex, real, and non-negative real matrices, respectively. The circularly symmetric complex Gaussian (CSCG) distribution with zero mean and covariance $\sigma^2 \mathbf{I}$ is represented as $\mathcal{CN}(\mathbf{0}, \sigma^2 \mathbf{I})$, and the uniform distribution over $[a, b]$ is denoted by $\mathcal{U}[a, b]$. Operator $\partial\left(\cdot\right)$ denotes the partial derivative. $\nabla_{\mathbf{x}}f\left(\mathbf{x}\right)$ and $\nabla_{\mathbf{x}}^2f\left(\mathbf{x}\right)$ denote the gradient vector and Hessian matrix of $f\left(\mathbf{x}\right)$ w.r.t. $\mathbf{x}$ respectively.

\section{Related Work}\label{sec:related}
\begin{figure*}[!t]
    \centering
    \includegraphics[width=0.95\textwidth]{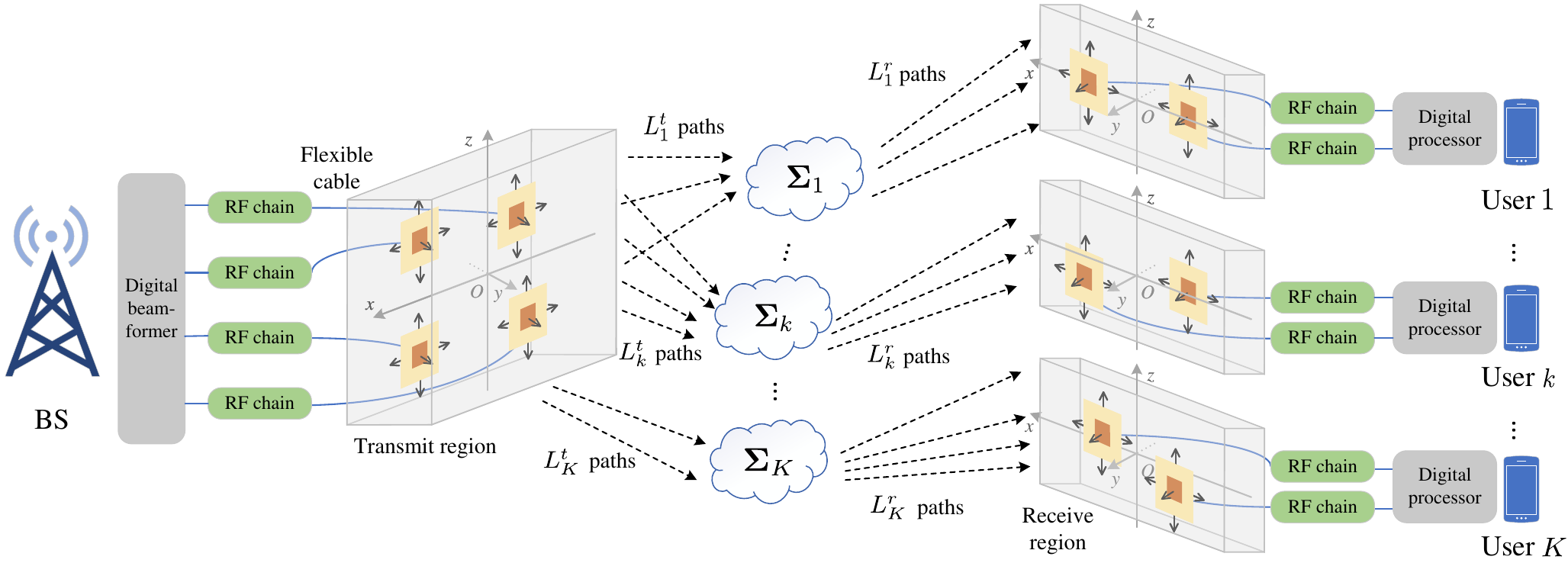}
    \caption{The 3D FA-assisted MU-MIMO wireless communication network.}
    \label{fig:system}
    \vspace{-0.5\baselineskip}
\end{figure*}
The first FA position optimization algorithm was introduced in~\cite{maMIMOCapacityCharacterization2024}, where the authors considered a point-to-point MIMO system. They formulated a non-convex optimization problem and proposed an SCA-based algorithm to solve the problem. However, this work did not consider joint beamforming and FA position optimization. Due to the strong coupling between beamforming matrices and FA positions, the SCA-based approach is not directly applicable to the joint optimization problem. To address this coupling, references~\cite{zhuMovableAntennaEnhancedMultiuser2024,huFluidAntennasEnabledMultiuser2024} investigated joint beamforming and FA position optimization in uplink MU-MISO systems. Both works employed ZF and/or MMSE techniques to decouple beamforming and FA positions, followed by FA position optimization using multi-directional descent (MDD) and projected gradient descent (PGD), respectively. However, ZF and MMSE can lead to significant performance degradation due to their oversimplification of the objective function, and the convergence of MDD and PGD is not guaranteed. An alternative approach leverages swarm intelligence algorithms such as particle swarm optimization (PSO)~\cite{xiaoMultiuserCommunicationsMovableAntenna2024}, which offers global search capabilities and achieves better performance than MDD- and PGD-based methods. Nevertheless, PSO incurs substantially higher computational complexity, limiting its practicality for large-scale systems.

To better balance performance and complexity, reference~\cite{tangSecureMIMOCommunication2025} considered a multiple-input single-output single-eavesdropper (MISOSE) system and formulated secure rate maximization as a non-convex optimization problem. The authors combined the MMSE-based decoupling method with a majorization–minimization (MM) algorithm for FA position optimization. However, the overall system performance remained constrained by the MMSE method. More recently, reference~\cite{fengWeightedSumRateMaximization2024} investigated the WSR maximization problem in an FA-assisted downlink MU-MISO system. The authors first applied the \emph{scalar} WMMSE algorithm to decouple the beamforming vectors and FA positions, and then adopted the MM framework to address the resulting non-convex FA position optimization problem.

In this paper, we considered the WSR maximization problem in the more general MU-MIMO setup. Although the \emph{scalar} WMMSE method~\cite{fengWeightedSumRateMaximization2024} effectively decouples beamforming and FA position optimization with satisfactory performance in MU-MISO systems, extending this technique to MU-MIMO systems is non-trivial~\cite{shenOptimizationMIMODevicetoDevice2019}. Instead, we adopt \emph{matrix} FP techniques to handle the decoupling in the MU-MIMO setting. For FA position optimization, all surrogate functions derived in SCA- or MM-based methods~\cite{maMIMOCapacityCharacterization2024,tangSecureMIMOCommunication2025,fengWeightedSumRateMaximization2024} are constructed only w.r.t. a single transmit or receive FA position. These methods update one FA position at a time while keeping the others fixed, which we refer to as \emph{sequential} SCA/MM. Such sequential updates result in lower computational efficiency\footnote{The high computational cost of the sequential MM algorithm in~\cite{fengWeightedSumRateMaximization2024} is due to the loop-based iterative updates. In contrast, the parallel MM algorithm proposed in this paper can be efficiently implemented using matrix operations.} and is unsuitable for decentralized implementations. By contrast, the proposed \emph{parallel} MM algorithm constructs a surrogate function jointly w.r.t. all transmit or receive FAs and updates them simultaneously.

\section{System Model and Problem Formulation}\label{sec:system}
As shown in Fig.~\ref{fig:system}, we consider a downlink MU-MIMO system where a base station (BS) with $M$ FAs serves $K$ users, each equipped with $N$ FAs. A three-dimensional (3D) Cartesian coordinate system is established to describe the positions of the transmit FAs at the BS and receive FAs at the users. Specifically, let $\mathbf{t}_m =\left[x_m^\mathtt{Tx},y_m^\mathtt{Tx},z_m^\mathtt{Tx}\right]^\transpose\in\mathcal{C}_m^\mathtt{Tx}$, $1\leq m\leq M$, denote the position of $m$-th transmit FA at the BS and let $\mathbf{r}_{kn} =\left[x_{kn}^\mathtt{Rx},y_{kn}^\mathtt{Rx},z_{kn}^\mathtt{Rx}\right]^\transpose\in\mathcal{C}_{kn}^\mathtt{Rx}$, $1\leq k\leq K$, $1\leq n\leq N$, denote the position of the $n$-th receive FA at user $k$, where $\mathcal{C}_m^\mathtt{Tx}$ and $\mathcal{C}_{kn}^\mathtt{Rx}$ are the given 3D movable regions of transmit and receive FAs. Without loss of generality, the movable regions are assumed to be cuboid~\cite{zhuMovableAntennaEnhancedMultiuser2024}, i.e., $\mathcal{C}_m^\mathtt{Tx}=\left[x_m^{\min},x_m^{\max}\right]\times \left[y_m^{\min},y_m^{\max}\right]\times \left[z_m^{\min},z_m^{\max}\right]$ and $\mathcal{C}_{kn}^\mathtt{Rx}=\left[x_{kn}^{\min},x_{kn}^{\max}\right]\times \left[y_{kn}^{\min},y_{kn}^{\max}\right]\times \left[z_{kn}^{\min},z_{kn}^{\max}\right]$, $\forall k,n$, where we assume the FA movable regions of the $K$ users to be identical. In this paper, we make the following assumptions.
\begin{itemize}
    \item \textit{Narrow-band slow fading:} The transmit and receive FAs remain static or move slowly within the movable region during each quasi-static fading block; The time required for FA movement is assumed to be much shorter than the coherence time~\cite{maMIMOCapacityCharacterization2024}.
    \item \textit{Knowledge of perfect channel state information (CSI):} The BS is assumed to have perfect knowledge of the downlink CSI for all users. CSI acquisition in FAS can be achieved using the methods proposed in~\cite{maCompressedSensingBased2023,xiaoChannelEstimationMovable2024,zhangChannelEstimationMovableAntenna2024,tangAccurateFastChannel2025}.
    \item \textit{Far-field condition:} The movable regions at both the users and the BS are much smaller than the signal propagation distance. The angle of departure (AoD), the angle of arrival (AoA), and the amplitude of the complex gain for each channel path are invariant to the FA positions~\cite{zhuModelingPerformanceAnalysis2024}.
    \item \textit{Continuous FA movement:} The FA positions are assumed to be continuous and can be adjusted to any location within the movable region.\footnote{Due to the limited precision of stepper motors and baseband processors, motor-driven FASs involve discrete optimization at certain stages. Nevertheless, their precision is significantly higher than that of pixel-based FASs. As a result, most existing works that aim to evaluate performance limits~\cite{ma2023capacity,zhuMovableAntennaEnhancedMultiuser2024,tangSecureMIMOCommunication2025,fengWeightedSumRateMaximization2024} relax the discrete optimization problem to a continuous one. We adopt the same assumption in this work.}
\end{itemize}

\subsection{Signal Model}\label{subsec:signal}
Let $\mathbf{s}_k \in \mathbb{C}^{d}$ denote the data stream intended for user $k$, where $d \leq \min\{M, N\}$ represents the number of parallel data streams. Assume that $\mathbf{s}_k \sim \mathcal{CN}(\mathbf{0}, \mathbf{I})$. Define $\mathbf{W}_k\in\mathbb{C}^{M\times d}$ as the beamforming matrix for transmitting data $\mathbf{s}_k$ from the BS to user $k$. The received signal $\mathbf{y}_k$ at user $k$ is given by\footnote{In this paper, we do not consider the digital combining matrix. But both the proposed centralized and decentralized algorithms can be easily extended to the case with digital combiners.}
\begin{small}
\begin{equation}\label{eq:rx_signal}
 \mathbf{y}_k=\mathbf{H}_k\left(\mathbf{T},\mathbf{R}_k\right)\mathbf{W}_k\mathbf{s}_k+\sum_{j=1, j\neq k}^{K}\mathbf{H}_k\left(\mathbf{T},\mathbf{R}_k\right)\mathbf{W}_j\mathbf{s}_j+\mathbf{n}_k,
\end{equation}
\end{small}%
where $\mathbf{T}=[\mathbf{t}_1, \dots, \mathbf{t}_M]^\transpose \in \mathbb{R}^{M\times 3}$ and $\mathbf{R}_k=[\mathbf{r}_{k1}, \dots, \mathbf{r}_{kN}]^\transpose \in \mathbb{R}^{N\times 3}$ represent the positions of the transmit FAs at the BS and the receive FAs at user $k$, respectively. The channel matrix between the BS and user $k$ is given as $\mathbf{H}_k\left(\mathbf{T},\mathbf{R}_k\right)\in\mathbb{C}^{N\times M}$, which depends on both $\mathbf{T}$ and $\mathbf{R}_k$. The term $\mathbf{n}_k\in\mathbb{C}^{N}$ denotes additive white Gaussian noise following the distribution $\mathcal{CN}\left(\mathbf{0},\sigma_k^2\mathbf{I}\right)$.

\subsection{Channel Model}\label{subsec:channel}
Let $L_k^\mathtt{Tx}$ and $L_k^\mathtt{Rx}$ denote the numbers of transmit and receive channel paths between the BS and user $k$, respectively. The direction vectors corresponding to the $q$-th transmit and receive paths are given by
\begin{align}\label{eq:steering_vector}
 \mathbf{g}_{kq}^\mathtt{Tx}&=\left[\cos\theta_{kq}^\mathtt{Tx}\cos\phi_{kq}^\mathtt{Tx}, \cos\theta_{kq}^\mathtt{Tx}\sin\phi_{kq}^\mathtt{Tx}, \sin\theta_{kq}^\mathtt{Tx}\right]^{\transpose},\\
 \mathbf{f}_{kq}^\mathtt{Rx}&=\left[\cos\theta_{kq}^\mathtt{Rx}\cos\phi_{kq}^\mathtt{Rx},\cos\theta_{kq}^\mathtt{Rx}\sin\phi_{kq}^\mathtt{Rx},\sin\theta_{kq}^\mathtt{Rx}\right]^{\transpose},
\end{align}
where $\theta_{kq}^\mathtt{Tx}$ and $\phi_{kq}^\mathtt{Tx}$ (and $\theta_{kq}^\mathtt{Rx}$ and $\phi_{kq}^\mathtt{Rx}$) are the elevation and azimuth AoDs (and AoAs) of the $q$-th path between the BS and user $k$. For the $q$-th transmit (and receive) channel path from the BS to user $k$, the distance difference between the path originating from the $m$-th BS antenna position $\mathbf{t}_m$ (and $n$-th user antenna position $\mathbf{r}_{kn}$) and that from the origin of the BS (and user $k$) coordinate system are given by
\begin{equation}\label{eq:distance_diff}
    \rho_{kq}^\mathtt{Tx}(\mathbf{t}_m) \triangleq \left(\mathbf{g}_{kq}^\mathtt{Tx}\right)^\transpose \mathbf{t}_m, \qquad
    \rho_{kq}^\mathtt{Rx}(\mathbf{r}_{kn}) \triangleq \left(\mathbf{f}_{kq}^\mathtt{Rx}\right)^\transpose \mathbf{r}_{kn},
\end{equation}
respectively. The transmit and receive field-response vectors (FRVs) between the BS and user $k$ are given by~\cite{zhuMovableAntennaEnhancedMultiuser2024}
\begin{align}
 \mathbf{g}_k(\mathbf{t}_m)&\triangleq\big[\mathrm{e}^{\jmath\frac{2\pi}{\lambda}\rho_{k1}^\mathtt{Tx}(\mathbf{t}_m)},\dots, \mathrm{e}^{\jmath\frac{2\pi}{\lambda}\rho_{k,L_k^\mathtt{Tx}}^\mathtt{Tx}(\mathbf{t}_m)}\big]^{\transpose},\label{eq:tx_FRV}\\
 \mathbf{f}_k(\mathbf{r}_{kn})&\triangleq\big[\mathrm{e}^{\jmath\frac{2\pi}{\lambda}\rho_{k1}^\mathtt{Rx}(\mathbf{r}_{kn})},\cdots, \mathrm{e}^{\jmath\frac{2\pi}{\lambda}\rho_{k,L_k^\mathtt{Rx}}^\mathtt{Rx}(\mathbf{r}_{kn})}\big]^{\transpose},\label{eq:rx_FRV}
\end{align} 
respectively, where $\lambda$ denotes the carrier wavelength. By defining the path-response matrix (PRM) $\mathbf{\Sigma}_k\in\mathbb{C}^{L_k^{\mathtt{Rx}}\times L_k^{\mathtt{Tx}}}$ as the response between each pair of transmit and receive channel paths from the BS to user $k$, the channel matrix $\mathbf{H}_k\left(\mathbf{T},\mathbf{R}_k\right)$ is given by
\begin{equation}\label{eq:channel_k}
 \mathbf{H}_k\left(\mathbf{T}, \mathbf{R}_k\right) = \mathbf{F}_k^{\hermconj}\left(\mathbf{\mathbf{R}_k}\right)\mathbf{\Sigma}_k\mathbf{G}_k\left(\mathbf{T}\right),
\end{equation}
where $\mathbf{F}_k\left(\mathbf{R}_k\right)=\left[\mathbf{f}_k(\mathbf{r}_{k1}),\dots,\mathbf{f}_k(\mathbf{r}_{kn})\right]$ and $\mathbf{G}_k\left(\mathbf{T}\right)=\left[\mathbf{g}_k(\mathbf{t}_1),\dots,\mathbf{g}_k(\mathbf{t}_M)\right]$ denote the field response matrices (FRMs) of all the receive FAs at user $k$ and those at the BS, respectively.

\subsection{Problem Formulation}\label{subsec:problem}
A fundamental problem in MU-MIMO downlink transmission is WSR maximization. The WSR is defined as  
\begin{equation}\label{eq:def_wsr}
 R=\sum\nolimits_{k=1}^{K}\alpha_k R_k,
\end{equation}
where the weight $\alpha_k$ denotes the priority of user $k$, and $R_k$ is the achievable rate of user $k$, given by~\cite{vishwanathCapacityMultipleInput2002,stankovicGeneralizedDesignMultiUser2008}
\vspace{0.2cm}
\begin{equation}\label{eq:achievable_rate}
 R_k = \log\det\left(\mathbf{I}+\mathbf{W}_k^{\hermconj}\mathbf{H}_k^{\hermconj}\left(\mathbf{T},\mathbf{R}_k\right)\mathbf{M}_k^{-1}\mathbf{H}_k\left(\mathbf{T},\mathbf{R}_k\right)\mathbf{W}_k\right).
\end{equation}
The interference-plus-noise matrix $\mathbf{M}_k$ is defined as
\begin{equation}\label{eq:snr_matrix}
 \mathbf{M}_k=\sum_{j=1,j\neq k}^{K}\mathbf{H}_k\left(\mathbf{T},\mathbf{R}_k\right)\mathbf{W}_j\mathbf{W}_j^{\hermconj}\mathbf{H}_k^{\hermconj}\left(\mathbf{T},\mathbf{R}_k\right)+\sigma_k^2\mathbf{I}.
\end{equation}
Let $\underline{\mathbf{W}}=\{\mathbf{W}_k,\forall k\}$ denote the set of beamforming matrices, and $\underline{\mathbf{R}}=\{\mathbf{R}_k,\forall k\}$ denote the set of all receive FA positions. Then, we can formulate the optimization problem as
\begin{figure*}[hb]
    \normalsize
    \vspace{-0.5\baselineskip}
    \hrulefill
    \setcounter{MYtempeqncnt}{\value{equation}}
    \setcounter{equation}{12}
    \begin{small}
    \begin{equation}\label{eq:lagrangian_dual_trans}
        \begin{aligned}
 f_{\rm Lag}\left(\underline{\mathbf{W}}, \mathbf{T}, \underline{\mathbf{R}}, \underline{\mathbf{\Gamma}}\right)=
            &\sum\nolimits_{k=1}^{K}\alpha_k\left(\log\det\left(\mathbf{I}+\mathbf{\Gamma}_k\right)-\trace\left(\mathbf{\Gamma}_k\right)+\trace\left(\left(\mathbf{I}+\mathbf{\Gamma}_k\right)\mathbf{W}_k^{\hermconj}\mathbf{H}_k^{\hermconj}\left(\mathbf{T},\mathbf{R}_k\right)\right.\right.\\
            &\left.\left.\times\left(\sum\nolimits_{j=1}^{K}\mathbf{H}_k\left(\mathbf{T},\mathbf{R}_k\right)\mathbf{W}_j\mathbf{W}_j^{\hermconj}\mathbf{H}_k^{\hermconj}\left(\mathbf{T},\mathbf{R}_k\right)+\sigma_k^2\mathbf{I}\right)^{-1}\mathbf{H}_k\left(\mathbf{T},\mathbf{R}_k\right)\mathbf{W}_k\right)\right).
        \end{aligned}
    \end{equation}
    \end{small}%
    \setcounter{equation}{\value{MYtempeqncnt}}
\end{figure*}
\begin{figure*}[hb]
    \normalsize
    \vspace{-0.5\baselineskip}
    \hrulefill
    \setcounter{MYtempeqncnt}{\value{equation}}
    \setcounter{equation}{14}
    \begin{small}
    \begin{equation}\label{eq:quadratic_trans}
        \begin{aligned}
 f_{\rm Quad}\left(\underline{\mathbf{W}}, \mathbf{T}, \underline{\mathbf{R}}, \underline{\mathbf{\Gamma}}, \underline{\mathbf{\Phi}}\right)=
            &\sum\nolimits_{k=1}^{K}\left(\alpha_k\log\det\left(\mathbf{I}+\mathbf{\Gamma}_k\right)-\alpha_k\trace\left(\mathbf{\Gamma}_k\right) +\trace\left(\left(\mathbf{I}+\mathbf{\Gamma}_k\right)\left(\sqrt{\alpha_k}\mathbf{W}_k^{\hermconj}\mathbf{H}_{k}^{\hermconj}\left(\mathbf{T},\mathbf{R}_k\right)\mathbf{\Phi}_k \right.\right.\right.\\
            &\left.\left.\left.+ \sqrt{\alpha_k}\mathbf{\Phi}_k^{\hermconj}\mathbf{H}_k\left(\mathbf{T},\mathbf{R}_k\right) \mathbf{W}_k-\mathbf{\Phi}_k^{\hermconj}\left(\sum\nolimits_{j=1}^{K}\mathbf{H}_k\left(\mathbf{T},\mathbf{R}_k\right)\mathbf{W}_j\mathbf{W}_j^{\hermconj}\mathbf{H}_k^{\hermconj}\left(\mathbf{T},\mathbf{R}_k\right)+\sigma_k^2\mathbf{I}\right)\mathbf{\Phi}_k\right)\right)\right).
        \end{aligned}
    \end{equation}
    \end{small}%
    \setcounter{equation}{\value{MYtempeqncnt}}
\end{figure*}
\begin{subequations}\label{opt:problem}
    \begin{align}              
 \underset{\underline{\mathbf{W}}, \mathbf{T},\underline{\mathbf{R}}}{\max}\ \ &R\label{opt:obj}\\
 {\text{s. t.}}\ \ \ &\sum\nolimits_{k=1}^{K}\trace\left(\mathbf{W}_k\mathbf{W}_k^{\hermconj}\right)\leq P_{\max},\label{opt:power_constraint}\\
        &\mathbf{t}_m\in\mathcal{C}_m^\mathtt{Tx},\ \forall m,\label{opt:tx_antenna_constraint}\\ 
        &\mathbf{r}_{kn}\in\mathcal{C}_{kn}^\mathtt{Rx},\ \forall{kn},\label{opt:rx_antenna_constraint}\\
        &\|\mathbf{t}_m-\mathbf{t}_{m^\prime}\|_2\geq D,\ 1\leq m, m^\prime\leq M, m\neq m^\prime,\label{opt:tx_coupling_constraint}\\
        &\|\mathbf{r}_{kn}-\mathbf{r}_{kn^\prime}\|_2\geq D,\ \forall k,\ 1\leq n, n^\prime\leq N, n\neq n^\prime.\label{opt:rx_coupling_constraint}
    \end{align}
\end{subequations} 
Here, the constraint~\eqref{opt:power_constraint} denotes the total transmit power constraint, where $P_{\max}$ is the total transmit power budget at the BS. Constraints~\eqref{opt:tx_antenna_constraint} and~\eqref{opt:rx_antenna_constraint} guarantee the FAs at the BS and users remain within the movable regions. Constraints~\eqref{opt:tx_coupling_constraint} and~\eqref{opt:rx_coupling_constraint} prevent mechanical collision between any pair of FAs at the BS and users, respectively. The problem~\eqref{opt:problem} is difficult to solve because the objective function~\eqref{opt:obj} is highly non-linear and non-concave w.r.t. the beamforming matrices $\underline{\mathbf{W}}$ and the FA positions $\mathbf{T}$ and $\underline{\mathbf{R}}$. Additionally, the optimization variables are highly coupled, making the problem more intractable. 

\section{Block Coordinate Ascent (BCA)-Based Algorithm}\label{sec:bca_sol}
In this section, we propose a BCA-based algorithm for the problem~\eqref{opt:problem}. First, we employ the matrix FP method to decouple the variables in the problem~\eqref{opt:problem}. Then, the MM algorithm is utilized to address the non-convex optimization of FA positions.

\subsection{Problem Reformulation}\label{subsec:problem_reform}
To solve the complicated problem~\eqref{opt:problem}, we first reformulate it using the matrix FP method~\cite{shen2018fractional}. Specifically, since the objective function~\eqref{opt:obj} is a sum-of-functions-of-matrix-ratios, we apply the matrix FP framework developed in~\cite{shenOptimizationMIMODevicetoDevice2019}, as detailed below.

First, applying the matrix Lagrangian dual transform~\cite[Theorem~2]{shenOptimizationMIMODevicetoDevice2019} to the problem~\eqref{opt:problem} allows us to extract the ratios from the logarithms in~\eqref{eq:achievable_rate}. Therefore, the problem~\eqref{opt:problem} can be reformulated as
\begin{subequations}\label{opt:problem_lagrangian_dual}
    \begin{align}              
 \underset{\underline{\mathbf{W}}, \mathbf{T}, \underline{\mathbf{R}}, \underline{\mathbf{\Gamma}}}{\max}\ \ &f_{\rm Lag}\left(\underline{\mathbf{W}}, \mathbf{T}, \underline{\mathbf{R}}, \underline{\mathbf{\Gamma}}\right)\label{opt:obj_lagrangian_dual}\\
 {\text{s. t.}}\ \ \ \ &\eqref{opt:power_constraint}-\eqref{opt:rx_coupling_constraint},
    \end{align}
\end{subequations} 
where $f_{\rm Lag}\left(\underline{\mathbf{W}}, \mathbf{T}, \underline{\mathbf{R}}, \underline{\mathbf{\Gamma}}\right)$ is given by~\eqref{eq:lagrangian_dual_trans} at the bottom of the page, and $\underline{\mathbf{\Gamma}}=\{\mathbf{\Gamma}_k,\forall k\}$ denotes the set of auxiliary variables.

\addtocounter{equation}{1}

By applying the matrix quadratic transform~\cite[Theorem~1]{shenOptimizationMIMODevicetoDevice2019} to the problem~\eqref{opt:problem_lagrangian_dual}, we can further decouple the ratios in the reformulated objective function~\eqref{eq:lagrangian_dual_trans} and reformulate the problem~\eqref{opt:problem_lagrangian_dual} as
\begin{subequations}\label{opt:problem_quadratic}
    \begin{align}              
 \underset{\underline{\mathbf{W}}, \mathbf{T}, \underline{\mathbf{R}}, \underline{\mathbf{\Gamma}}, \underline{\mathbf{\Phi}}}{\max}\ \ &f_{\rm Quad}\left(\underline{\mathbf{W}}, \mathbf{T}, \underline{\mathbf{R}}, \underline{\mathbf{\Gamma}}, \underline{\mathbf{\Phi}}\right)\label{opt:obj_quadratic}\\
 {\text{s. t.}}\ \ \ \ \ &\eqref{opt:power_constraint}-\eqref{opt:rx_coupling_constraint},
    \end{align}
\end{subequations} 
where $f_{\rm Quad}\left(\underline{\mathbf{W}}, \mathbf{T}, \underline{\mathbf{R}}, \underline{\mathbf{\Gamma}}, \underline{\mathbf{\Phi}}\right)$ is given by~\eqref{eq:quadratic_trans} at the bottom of the page, and $\underline{\mathbf{\Phi}}=\{\mathbf{\Phi}_k,\forall k\}$ denotes the set of auxiliary variables. 

\addtocounter{equation}{1}

With the above FP-based two-step transformation, the original problem~\eqref{opt:problem} is equivalent to the problem~\eqref{opt:problem_quadratic}. Then we employ the BCA algorithm to solve the problem~\eqref{opt:problem_quadratic} by iteratively optimizing one set of variables while keeping others fixed until convergence.

\subsection{Update Step of $\underline{\mathbf{\Gamma}}$ and $\underline{\mathbf{\Phi}}$}
In this step, we aim to optimize the auxiliary variables $\underline{\mathbf{\Gamma}}$ and $\underline{\mathbf{\Phi}}$ with the fixed $\underline{\mathbf{W}}$, $\mathbf{T}$, and $\underline{\mathbf{R}}$. The optimal solution of $\underline{\mathbf{\Gamma}}$ and $\underline{\mathbf{\Phi}}$ are derived by setting the first-order derivatives of~\eqref{eq:quadratic_trans} to zero w.r.t. $\mathbf{\Gamma}_k$ and $\mathbf{\Phi}_k$, respectively. Let $\overline{{\mathbf{W}}}_k$, $\overline{\mathbf{T}}$, $\overline{\mathbf{R}}_k$, $\overline{\mathbf{\Gamma}}_k$, and $\overline{\mathbf{\Phi}}_k$ denote the temporal optimization results obtained by the previous iteration, and let $\overline{\mathbf{M}}_k$ denote the temporal interference-plus-noise matrix. The closed-form expressions for the optimal $\mathbf{\Phi}_k$ and $\mathbf{\Gamma}_k$ are given by~\cite{shenOptimizationMIMODevicetoDevice2019}
\begin{align}
 \mathbf{\Phi}_k &= \sqrt{\alpha_k}\left(\overline{\mathbf{M}}_k+\overline{\mathbf{H}}_k\overline{\mathbf{W}}_k\overline{\mathbf{W}}_k^{\hermconj}\overline{\mathbf{H}}_k^{\hermconj}\right)^{-1}\overline{\mathbf{H}}_{k}\overline{\mathbf{W}}_k,\label{eq:opt_phi}\\
 \mathbf{\Gamma}_k &= \overline{\mathbf{W}}_k^{\hermconj}\overline{\mathbf{H}}_k^{\hermconj}\overline{\mathbf{M}}_k^{-1}\overline{\mathbf{H}}_k\overline{\mathbf{W}}_k,\label{eq:opt_gamma}
\end{align}
respectively, where we define $\overline{\mathbf{H}}_k\triangleq\mathbf{H}_k\left(\overline{\mathbf{T}}, \overline{\mathbf{R}}_k\right)$.

\subsection{Update Step of $\underline{\mathbf{W}}$}\label{subsec:bisec_beamforming}
In this step, we aim to optimize the beamforming matrices $\underline{\mathbf{W}}$ with the fixed $\mathbf{T}$, $\underline{\mathbf{R}}$, $\underline{\mathbf{\Gamma}}$, and $\underline{\mathbf{\Phi}}$. Then, the optimization problem~\eqref{opt:problem_quadratic} reduces to
\begin{equation}\label{opt_beamforming:problem}        
 \underset{\underline{\mathbf{W}}}{\max}\ f_{\rm Quad}\left(\underline{\mathbf{W}}\right)\qquad {\text{s. t.}}\ \eqref{opt:power_constraint},
\end{equation} 
where $f_{\rm Quad}\left(\underline{\mathbf{W}}\right)=f_{\rm Quad}\left(\underline{\mathbf{W}}, \overline{\mathbf{T}}, \overline{\underline{\mathbf{R}}}, \overline{\underline{\mathbf{\Gamma}}}, \overline{\underline{\mathbf{\Phi}}}\right)$. As derived in~\cite{shenOptimizationMIMODevicetoDevice2019}, the optimal solution to the problem~\eqref{opt_beamforming:problem} is given by
\begin{small}
\begin{equation}\label{eq:opt_beamforming}
 \mathbf{W}_k=\Big[\sum_{j=1}^{K}\sqrt{\alpha_k}\overline{\mathbf{H}}_j^{\hermconj}\overline{\mathbf{\Phi}}_j\left(\mathbf{I}+\overline{\mathbf{\Gamma}}_j\right)\overline{\mathbf{\Phi}}_j^{\hermconj}\overline{\mathbf{H}}_j+\mu\mathbf{I}\Big]^{-1} \overline{\mathbf{H}}_k^{\hermconj}\overline{\mathbf{\Phi}}_k\left(\mathbf{I}+\overline{\mathbf{\Gamma}}_k\right),
\end{equation}
\end{small}%
where $\mu\geq 0$ is computed via bisection search to ensure that $\overline{\mathbf{W}}$ can satisfy the complementary slackness condition of the power budget constraint~\eqref{opt:power_constraint}~\cite{boyd2004convex}.

\subsection{Update Step of $\mathbf{T}$}\label{subsec:opt_tx_position}
In this step, we aim to optimize the positions of the transmit FAs at the BS $\mathbf{T}$ with the fixed $\underline{\mathbf{W}}$, $\underline{\mathbf{R}}$, $\underline{\mathbf{\Gamma}}$, and $\underline{\mathbf{\Phi}}$. Then, the optimization problem~\eqref{opt:obj_quadratic} reduces to
\begin{equation}\label{opt_tx_position:iteration}          
 \underset{\mathbf{T}}{\max}\ \ f_{\rm Quad}\left(\mathbf{T}\right)\qquad{\text{s. t.}}\ \eqref{opt:tx_antenna_constraint},~\eqref{opt:tx_coupling_constraint}.
\end{equation} 
Unlike the update steps for $\underline{\mathbf{\Gamma}}$, $\underline{\mathbf{\Phi}}$, and $\underline{\mathbf{W}}$, the objective function $f_{\rm Quad}\left(\mathbf{T}\right)$ remains non-concave w.r.t. $\mathbf{T}$. The MM algorithm~\cite{sunMajorizationMinimizationAlgorithmsSignal2017} effectively solves this non-convex problem by iteratively finding a series of concave lower bounds for the non-concave function $f_{\rm Quad}\left(\mathbf{T}\right)$, known as the \textit{surrogate function}. A key advantage of the MM algorithm is its convergence guarantee, which is discussed in Section~\ref{subsec:convergence}.

To solve the problem~\eqref{opt_tx_position:iteration} using the MM algorithm, the optimal value of $\mathbf{T}$ is computed iteratively. Each MM iteration consists of a \textit{majorization step}, followed by a \textit{maximization step}. In the \textit{majorization step}, we construct the \textit{surrogate function} such that  
\begin{equation}\label{eq:mm_surrogate}
 h^{\mathtt{Tx}}\left(\mathbf{T}\big\vert\overline{\mathbf{T}}\right) \leq f_{\rm Quad}\left(\mathbf{T}\right),
\end{equation}
where $\mathtt{Tx}$ indicates the update step is related to $\mathbf{T}$. The equality~\eqref{eq:mm_surrogate} holds when $\mathbf{T} = \overline{\mathbf{T}}$, i.e.,
\begin{equation}\label{eq:mm_equality}
 h^{\mathtt{Tx}}\left(\overline{\mathbf{T}}\big\vert\overline{\mathbf{T}}\right) = f_{\rm Quad}\left(\overline{\mathbf{T}}\right).
\end{equation}
In the \textit{maximization step}, we determine the optimal $\mathbf{T}$ subject to the given constraints by solving the following convex optimization problem:
\begin{equation}\label{opt_tx_position:problem_mm}
 \underset{\mathbf{T}}{\max}\ \ h^{\mathtt{Tx}}\left(\mathbf{T} \big\vert \overline{\mathbf{T}}\right),\qquad {\text{s. t.}}\ \eqref{opt:tx_antenna_constraint}, \eqref{opt:tx_coupling_constraint}.
\end{equation}

In each MM iteration, a distinct \textit{surrogate function} is constructed, based on which a new solution $\mathbf{T}$ is obtained by maximizing the resulting concave function. Then, the update step of $\mathbf{T}$ for the proposed MM algorithm is executed until convergence. The key to applying the MM algorithm is constructing a suitable surrogate function. Here, we introduce the following lemma to construct the surrogate function that locally approximates the objective function~\cite{sunMajorizationMinimizationAlgorithmsSignal2017}.

\begin{lemma}\label{lemma:taylor}
 Let $f: \mathbb{R}^n\to\mathbb{R}$ be a continuously differentiable function with bounded curvature, i.e., there exists a matrix $\mathbf{L}$ such that $\mathbf{L} \succeq \nabla^2 f(\mathbf{x}),\ \forall\mathbf{x} \in \mathcal{X}$. Then,  
    \begin{equation}\label{eq:taylor}
 f(\mathbf{x}) \geq f(\overline{\mathbf{x}}) + \nabla f^\transpose(\overline{\mathbf{x}}) \left(\mathbf{x} - \overline{\mathbf{x}}\right) - \frac12 \left(\mathbf{x} - \overline{\mathbf{x}}\right)^\transpose \mathbf{L} \left(\mathbf{x} - \overline{\mathbf{x}}\right),
    \end{equation}
 where $\overline{\mathbf{x}}$ is a constant satisfying $\overline{\mathbf{x}} \in \mathcal{X}$.    
\end{lemma}

We apply Lemma~\ref{lemma:taylor} to $f_{\rm Quad}\left(\mathbf{T}\right)$ and let $\mathbf{L} = \delta^{\mathtt{Tx}} \mathbf{I}$ with
\begin{equation}\label{eq:opt_tx_position_delta_bound}
    \delta^{\mathtt{Tx}} \geq \lambda_{\max}\left(\nabla_{\vect\left(\mathbf{T}\right)}^2 f_{\rm Quad}\left(\mathbf{\mathbf{T}}\right)\right),
\end{equation}
for any given $\mathbf{T}$ satisfying~\eqref{opt:tx_antenna_constraint} and~\eqref{opt:tx_coupling_constraint}, where $\lambda_{\max}\left(\cdot\right)$ denotes the maximum eigenvalue of a matrix. Then, we can construct the surrogate function $h^{\mathtt{Tx}}\left(\mathbf{T}\big\vert\overline{\mathbf{T}}\right)$ as
\begin{multline}\label{eq:opt_tx_position_surr}
 h^{\mathtt{Tx}}\left(\mathbf{T}\big\vert\overline{\mathbf{T}}\right) = -\frac{\delta^{\mathtt{Tx}}}2\vect\left(\mathbf{T}\right)^\transpose \vect\left(\mathbf{T}\right) \\
 + \left(\nabla_{\vect\left(\mathbf{T}\right)} f_{\rm Quad}^\transpose\left(\overline{\mathbf{T}}\right) + \delta^{\mathtt{Tx}}\vect\left(\overline{\mathbf{T}}\right)^\transpose\right)\vect\left(\mathbf{T}\right) + \mathrm{const}.
\end{multline}
Finding the surrogate function $h^{\mathtt{Tx}}\left(\mathbf{T}\big\vert\overline{\mathbf{T}}\right)$ is thus equivalent to computing the gradient $\nabla_{\vect\left(\mathbf{T}\right)} f_{\rm Quad}\left(\mathbf{T}\right)$ and determining the constant $\delta^{\mathtt{Tx}}$. Since $\mathbf{T}$ influences $f_{\rm Quad}\left(\mathbf{T}\right)$ only through the matrix $\mathbf{G}_k$ as defined in~\eqref{eq:quadratic_trans}, we compute the gradient $\nabla_{\vect\left(\mathbf{T}\right)} f_{\rm Quad}\left(\mathbf{T}\right)$ using the matrix chain rule. Let $\mathbf{D}_k^{\mathtt{Tx}}$ denote the transpose of the first-order derivative of $f_{\rm Quad}\left(\mathbf{T}\right)$ w.r.t. $\mathbf{G}_k$, and it is given by~\cite{hjorungnesComplexValuedMatrixDifferentiation2007}
\begin{align}\label{eq:opt_tx_position_D}
 \mathbf{D}_k^\mathtt{Tx} &\triangleq \left(\frac{\partial f_{\rm Quad}}{\partial \mathbf{G}_k}\right)^\transpose \nonumber \\
    &= \sqrt{\alpha_k} \overline{\mathbf{W}}_k \left(\mathbf{I} + \overline{\mathbf{\Gamma}}_k\right) \overline{\mathbf{\Phi}}_k^\hermconj \overline{\mathbf{F}}_k^\hermconj \mathbf{\Sigma}_k 
 - \hat{\mathbf{W}} \mathbf{G}_k^\hermconj\left(\overline{\mathbf{T}}\right) \hat{\mathbf{\Sigma}}_k,
\end{align}
where we define $\overline{\mathbf{F}}_k \triangleq \mathbf{F}_k\left(\overline{\mathbf{R}}_k\right)$, $\hat{\mathbf{W}} \triangleq \sum_{k=1}^K \overline{\mathbf{W}}_k \overline{\mathbf{W}}_k^\hermconj$, and $\hat{\mathbf{\Sigma}}_k^{\mathtt{Tx}} \triangleq \mathbf{\Sigma}_k^\hermconj \overline{\mathbf{F}}_k \overline{\mathbf{\Phi}}_k \left(\mathbf{I} + \overline{\mathbf{\Gamma}}_k\right) \overline{\mathbf{\Phi}}_k^\hermconj \overline{\mathbf{F}}_k^\hermconj \mathbf{\Sigma}_k$. Based on $\mathbf{D}_k^\mathtt{Tx}$, the entries of $\nabla_{\vect\left(\mathbf{T}\right)} f_{\rm Quad}\left(\mathbf{T}\right)$ can be computed as
\begin{align}\label{eq:opt_tx_position_dfdxyz}
 \frac{\partial f_{\rm Quad}}{\partial \mathbf{t}_m} &= \left[\frac{\partial f_{\rm Quad}}{\partial x_m}, \frac{\partial f_{\rm Quad}}{\partial y_m}, \frac{\partial f_{\rm Quad}}{\partial z_m}\right]^\transpose \nonumber \\
    &= -\frac{4\pi}{\lambda} \sum_{k=1}^K \sum_{q=1}^{L_k^\mathtt{Tx}} \Big\lvert[\mathbf{D}_k^\mathtt{Tx}]_{mq}\Big\rvert \sin(\xi_{kmq}^{\mathtt{Tx}}) \mathbf{g}_{kq}^\mathtt{Tx},
\end{align}
where $\xi_{kmq}^{\mathtt{Tx}}$ is calculated by
\begin{align}\label{eq:opt_tx_position_xi}
    \xi_{kmq}^{\mathtt{Tx}} = &\angle[\mathbf{D}_k^\mathtt{Tx}]_{mq} + \frac{2\pi}{\lambda}\rho_{kq}^{\mathtt{Tx}}(\mathbf{t}_m).
\end{align}
The detailed derivation of~\eqref{eq:opt_tx_position_dfdxyz} is shown in Appendix~\ref{appendix:dfdt}. According to~\eqref{eq:opt_tx_position_delta_bound}, the constant $\delta^{\mathtt{Tx}}$ can be chosen by finding the upper bound of the maximum eigenvalue of the Hessian matrix $\nabla_{\vect\left(\mathbf{T}\right)}^2 f_{\rm Quad}\left(\mathbf{T}\right)$. Since calculating the eigenvalues of the Hessian matrix is computationally expensive, we first leverage the matrix infinity norm to upper bound the maximum eigenvalue. Then, we calculate the upper bound of the matrix infinity norm for \emph{all} possible $\mathbf{T}$. The detailed derivations of $\delta^{\mathtt{Tx}}$ is provided in Appendix~\ref{appendix:delta} and the result is given by
\begin{small}
    \begin{align}\label{eq:opt_tx_position_delta}
        \delta^{\mathtt{Tx}} =& \underset{1\leq m\leq M}{\max}\frac{24\pi^2}{\lambda^2}\sum_{k=1}^K \left[\Big(\sum_{j=1}^M \Big\lvert[\hat{\mathbf{W}}]_{mj}\Big\rvert + \sqrt{M}\Big\lVert[\hat{\mathbf{W}}]_m\Big\rVert_2\Big) \lVert\hat{\mathbf{\Sigma}}_k^{\mathtt{Tx}}\rVert_2 \right. \nonumber \\
        &+ \left.\sqrt{\frac{\alpha_k}{L_k^\mathtt{Tx}}} \Big\lVert[\overline{\mathbf{W}}_k]_m\left(\mathbf{I}+\overline{\mathbf{\Gamma}}_k\right)\overline{\mathbf{\Phi}}_k^\hermconj\overline{\mathbf{F}}_k^\hermconj\mathbf{\Sigma}_k^\hermconj\Big\rVert_2 \right]L_k^\mathtt{Tx}.
    \end{align}
\end{small}%

Although we have found the concave surrogate function~\eqref{eq:opt_tx_position_surr}, the non-convex constraint~\eqref{opt:tx_coupling_constraint} still makes the optimization problem~\eqref{opt_tx_position:problem_mm} non-convex. To make the problem tractable, we apply the Cauchy-Schwarz inequality to the left-hand side (l.h.s.) of the constraint~\eqref{opt:tx_coupling_constraint} and obtain a lower bound of $\lVert\mathbf{t}_m - \mathbf{t}_{m^\prime}\rVert_2$, given by
\begin{equation}\label{eq:opt_tx_position_cauchy}
 \lVert\mathbf{t}_m - \mathbf{t}_{m^\prime}\rVert_2 \geq \frac{\left(\overline{\mathbf{t}}_m - \overline{\mathbf{t}}_{m^\prime}\right)^\transpose\left(\mathbf{t}_m - \mathbf{t}_{m^\prime}\right)}{\lVert\overline{\mathbf{t}}_m - \overline{\mathbf{t}}_{m^\prime}\rVert_2}, \ \forall m \neq m^\prime.
\end{equation}
The inequality~\eqref{eq:opt_tx_position_cauchy} indicates that if $\mathbf{T}$ satisfies the following constraint
\begin{equation}\label{eq:opt_tx_position_stronger_constraint}
 \frac{\left(\overline{\mathbf{t}}_m - \overline{\mathbf{t}}_{m^\prime}\right)^\transpose\left(\mathbf{t}_m - \mathbf{t}_{m^\prime}\right)}{\lVert\overline{\mathbf{t}}_m - \overline{\mathbf{t}}_{m^\prime}\rVert_2} \geq D, \ \forall m \neq m^\prime,
\end{equation}
then $\mathbf{T}$ also satisfies constraint~\eqref{opt:tx_coupling_constraint}. In other words, constraint~\eqref{eq:opt_tx_position_stronger_constraint} is a sufficient condition for constraint~\eqref{opt:tx_coupling_constraint}. Therefore, the problem~\eqref{opt_tx_position:iteration} is iteratively solved by
\begin{equation}\label{opt_tx_position:iteration_final}
 \underset{\mathbf{T}}{\max}\ \ h^{\mathtt{Tx}}\left(\mathbf{T} \big\vert \overline{\mathbf{T}}\right), \qquad\\
 {\text{s. t.}}\ \eqref{opt:tx_antenna_constraint},\ \eqref{eq:opt_tx_position_stronger_constraint},
\end{equation}
which is a convex quadratic programming problem. As demonstrated in~\cite{maMIMOCapacityCharacterization2024}, solving~\eqref{opt_tx_position:iteration_final} can be simplified by initially assuming all constraints are inactive. This transformation reduces the problem to an unconstrained quadratic optimization, whose closed-form solution is given by 
\begin{equation}\label{eq:opt_tx_position_closed_form}
 \mathbf{T}^\star = \overline{\mathbf{T}} + \frac{1}{\delta^{\mathtt{Tx}}} \nabla_{\mathbf{T}} f_{\rm Quad}\left(\overline{\mathbf{T}}\right).
\end{equation}
Next, we verify this assumption by checking whether $\mathbf{T}^\star$ satisfies the constraints~\eqref{opt:tx_antenna_constraint} and~\eqref{eq:opt_tx_position_stronger_constraint}. If the constraints are not satisfied, we apply the interior-point method~\cite{boyd2004convex} to obtain a valid optimum $\mathbf{T}^\star$.

\textit{Remark:} Different from~\cite{ma2023capacity,fengWeightedSumRateMaximization2024}, the surrogate function $h^{\mathtt{Tx}}(\mathbf{T} \vert \overline{\mathbf{T}})$ in this paper is constructed w.r.t. all transmit FAs $\mathbf{T}$, enabling the \emph{parallel} update of all FA positions. Moreover, prior works~\cite{ma2023capacity,fengWeightedSumRateMaximization2024} employ the inequality from~\cite[Eq.~(26)]{sunMajorizationMinimizationAlgorithmsSignal2017} to eliminate the second-order term w.r.t. FA positions. In contrast, we construct the surrogate function directly from the gradient and Hessian of $f_{\rm Quad}$, yielding a tighter approximation and improved optimization performance. This construction relies on matrix chain rules~\eqref{eq:opt_position_chain} and the matrix infinity norm bound in~\eqref{eq:opt_tx_position_delta_ineq} detailed in Appendix~\ref{appendix:dfdt} and Appendix~\ref{appendix:delta}, respectively.

\subsection{Update Step of $\underline{\mathbf{R}}$}\label{subsec:opt_rx_position}
\begin{figure*}[hb]
    \normalsize
    \vspace{-\baselineskip}
    \hrulefill
    \setcounter{MYtempeqncnt}{\value{equation}}
    \setcounter{equation}{42}
    \begin{small}
    \begin{align}\label{eq:opt_rx_position_delta}
        \delta^{\mathtt{Rx}} = \underset{1\leq n\leq N}{\max}\frac{24\pi^2}{\lambda^2} L_k^\mathtt{Rx}\left[\Big(\sum_{j=1}^N\Big\lvert[\overline{\mathbf{\Phi}}_k]_n\left(\mathbf{I} + \overline{\mathbf{\Gamma}}_k\right)[\overline{\mathbf{\Phi}}_k]_j^\hermconj\Big\rvert + \sqrt{N}\Big\lVert[\overline{\mathbf{\Phi}}_k]_n\left(\mathbf{I} + \overline{\mathbf{\Gamma}}_k\right)\overline{\mathbf{\Phi}}_k^\hermconj\Big\rVert_2\Big)\lVert\hat{\mathbf{\Sigma}}_k^{\mathtt{Rx}}\rVert_2 + \sqrt{\frac{\alpha_k}{L_k^\mathtt{Rx}}}\Big\lVert[\overline{\mathbf{\Phi}}_k]_n\left(\mathbf{I} + \overline{\mathbf{\Gamma}}_k\right)\overline{\mathbf{W}}_k^\hermconj\overline{\mathbf{G}}_k^\hermconj\mathbf{\Sigma}_k^\hermconj\Big\rVert_2\right].
    \end{align}
    \end{small}
    \setcounter{equation}{\value{MYtempeqncnt}}
\end{figure*}

In this step, our target is to optimize the positions of the receive FAs $\underline{\mathbf{R}}$ with fixed $\mathbf{T}$, $\underline{\mathbf{W}}$, $\underline{\mathbf{\Gamma}}$, and $\underline{\mathbf{\Phi}}$. The objective function $f_{\rm Quad}\left(\underline{\mathbf{R}}\right)$ is reformulated as
\begin{equation}\label{eq:opt_rx_position_sum}
 f_{\rm Quad}\left(\underline{\mathbf{R}}\right) = \sum\nolimits_{k=1}^K f_{\rm Quad}\left(\mathbf{R}_k\right).
\end{equation}
Since the terms of the right hand side (r.h.s.) of~\eqref{eq:opt_rx_position_sum} do not couple with each other, it is feasible to optimize $f_{\rm Quad}\left(\mathbf{R}_k\right)$ independently. Therefore, we simply provide the update step of $\mathbf{R}_k$ in the remainder of this subsection. The optimization problem is then reformulated as
\begin{equation}\label{opt_rx_position:problem} 
 \underset{\mathbf{R}_k}{\max}\ f_{\rm Quad}\left(\mathbf{R}_k\right)\qquad {\text{s. t.}}\ \eqref{opt:rx_antenna_constraint},~\eqref{opt:rx_coupling_constraint}.
\end{equation}

Similar to the optimization step of $\mathbf{T}$, the objective function $f_{\rm Quad}\left(\mathbf{R}_k\right)$ is non-concave w.r.t. $\mathbf{R}_k$. Hence, we utilize MM to optimize $\mathbf{R}_k$, and the problem~\eqref{opt_rx_position:problem} is reformulated as
\begin{subequations}
    \begin{align}\label{opt_rx_position:iteration}
 \underset{\mathbf{\underline{R}}}{\max}\ \ &h^{\mathtt{Rx}}_k\left(\mathbf{R}_k\big\vert\overline{\mathbf{R}}_k\right)\\
 {\text{s. t.}}\ \ &\eqref{opt:rx_antenna_constraint},\nonumber\\
        &\frac{\left(\overline{\mathbf{r}}_{kn} - \overline{\mathbf{r}}_{kn^\prime}\right)^\transpose\left(\mathbf{r}_{kn} - \mathbf{r}_{kn^\prime}\right)}{\lVert\overline{\mathbf{r}}_{kn} - \overline{\mathbf{r}}_{kn^\prime}\rVert_2} \geq D,\ \forall n\neq n^{\prime},\label{eq:opt_rx_position_stronger_constraint}
    \end{align}
\end{subequations}
where $\mathtt{Rx}$ indicates the update step relates to $\underline{\mathbf{R}}$. The function $h^{\mathtt{Rx}}_k\left(\mathbf{R}_k\big\vert\overline{\mathbf{R}}_k\right)$ is the \textit{surrogate function} of $f_{\rm Quad}\left(\mathbf{R}_k\right)$:
\begin{small}
\begin{multline}\label{eq:opt_rx_position_surr}
 h^{\mathtt{Rx}}_k\left(\mathbf{R}_k\big\vert\overline{\mathbf{R}}_k\right) = -\frac{\delta^{\mathtt{Rx}}_k}2\vect\left(\mathbf{R}_k\right)^\transpose \vect\left(\mathbf{R}_k\right) \\
 + \left(\nabla_{\vect\left(\mathbf{R}_k\right)} f_{\rm Quad}^\transpose \left(\overline{\mathbf{R}}_k\right)+ \delta^{\mathtt{Rx}}_k\vect\left(\overline{\mathbf{R}}_k\right)^\transpose\right)\vect\left(\mathbf{R}_k\right) + \mathrm{const},
\end{multline}
\end{small}%
where $\delta_k^{\mathtt{Rx}}$ needs to satisfy
\begin{equation}\label{eq:opt_rx_position_delta_bound}
    \delta^{\mathtt{Rx}}_k \geq \lambda_{\max}\left(\nabla_{\vect\left(\mathbf{R}_k\right)}^2 f_{\rm Quad}\left(\mathbf{\mathbf{R}_k}\right)\right),
\end{equation}
for any $\mathbf{R}_k$ satisfying~\eqref{opt:rx_antenna_constraint} and~\eqref{eq:opt_rx_position_stronger_constraint} according to Lemma~\ref{lemma:taylor}. The entries of the gradient $\nabla_{\vect\left(\mathbf{R}_k\right)} f_{\rm Quad}\left(\mathbf{R}_k\right)$ are given by
\begin{subequations}\label{eq:opt_rx_position_dfdxyz}
    \begin{align}
 \frac{\partial f_{\rm Quad}}{\partial \mathbf{r}_{kn}} &= -\frac{4\pi}{\lambda}\sum_{k=1}^K\sum_{q=1}^{L_k^\mathtt{Rx}}\Big\lvert[\mathbf{D}_k^\mathtt{Rx}]_{nq}\Big\rvert\sin(\xi_{knq}^{\mathtt{Rx}})\mathbf{f}_{kq}^{\mathtt{Rx}}.
    \end{align}
\end{subequations}
The expressions of $\mathbf{D}_k^\mathtt{Rx}$ and $\xi_{knq}^{\mathtt{Rx}}$ are given by
\begin{small}
\begin{align}\label{eq:opt_rx_position_D}
 \mathbf{D}^{\mathtt{Rx}}_k = \left(\frac{\partial f_{\rm Quad}}{\partial \mathbf{F}_k}\right)^\transpose =& \sqrt{\alpha_k}\overline{\mathbf{\Phi}}_k\left(\mathbf{I} + \overline{\mathbf{\Gamma}}_k\right)\overline{\mathbf{W}}_k^\hermconj \overline{\mathbf{G}}_k^\hermconj \mathbf{\Sigma}_k^\hermconj \nonumber \\
    &- \overline{\mathbf{\Phi}}_k \left(\mathbf{I} + \overline{\mathbf{\Gamma}}_k\right) \overline{\mathbf{\Phi}}_k^\hermconj \mathbf{F}_k^\hermconj\left(\overline{\mathbf{R}}_k\right) \hat{\mathbf{\Sigma}}_k^{\mathtt{Rx}}
\end{align}
\end{small}%
and
\begin{equation}
    \xi_{knq}^{\mathtt{Rx}} = \angle[\mathbf{D}^{\mathtt{Rx}}_k]_{nq} + \frac{2\pi}{\lambda}\rho_{kq}^\mathtt{Rx}(\mathbf{r}_{kn}),
\end{equation}
respectively, where we define $\overline{\mathbf{G}}_k\triangleq\mathbf{G}_k\left(\overline{\mathbf{T}}\right)$ and $\hat{\mathbf{\Sigma}}_k^{\mathtt{Rx}} = \mathbf{\Sigma}_k \overline{\mathbf{G}}_k \hat{\mathbf{W}} \overline{\mathbf{G}}_k^\hermconj \mathbf{\Sigma}_k^\hermconj$. The closed-form expression of $\delta^{\mathtt{Rx}}_k$ satisfying~\eqref{eq:opt_rx_position_delta_bound} is given by~\eqref{eq:opt_rx_position_delta} at the bottom of the page. The derivations of~\eqref{eq:opt_rx_position_dfdxyz} and~\eqref{eq:opt_rx_position_delta} follow the same steps as~\eqref{eq:opt_tx_position_dfdxyz} and~\eqref{eq:opt_tx_position_delta}, respectively. Hence, we omit them for brevity.

\addtocounter{equation}{1}

The global optimal solution of $\mathbf{R}_k$ can be obtained in closed-form by assuming constraints~\eqref{opt:rx_antenna_constraint} and~\eqref{eq:opt_rx_position_stronger_constraint} are inactive, given by  
\begin{equation}\label{eq:opt_rx_position_closed_form}
 \mathbf{R}_k^\star = \overline{\mathbf{R}}_k + \frac{1}{\delta^{\mathtt{Rx}}_k} \nabla_{\mathbf{R}_k} f_{\rm Quad}\left(\overline{\mathbf{R}}_k\right).
\end{equation}
If $\mathbf{R}_k^\star$ does not satisfy constraint~\eqref{opt:rx_antenna_constraint} or~\eqref{eq:opt_rx_position_stronger_constraint}, we apply the interior-point method to obtain the optimal solution.

\subsection{Box-Constrained Movement Mode for FAs}\label{subsec:bca_box}
Although constraints~\eqref{eq:opt_tx_position_stronger_constraint} and~\eqref{eq:opt_rx_position_stronger_constraint} are linear and compatible with quadratic optimization algorithms, they introduce $\frac{1}{2} M(M-1)$ and $\frac{1}{2} NK(N-1)$ inequalities, respectively. As the number of constraints grows proportionally to $M^2$, solving problem~\eqref{opt:problem} becomes infeasible for large $M$. The original movement mode also presents challenges for practical implementation. Since all FAs share a common movable region, mechanical conflicts may arise, limiting the feasibility of the design in real-world applications~\cite{dongMovableAntennaWireless2024,ningMovableAntennaEnhancedWireless2024}. Moreover, under the DBP architecture discussed in Section~\ref{sec:dec_bca}, FAs from different clusters may violate the constraint~\eqref{opt:tx_coupling_constraint}, potentially leading to physical collisions.

Therefore, inspired by~\cite{maMIMOCapacityCharacterization2024,fengWeightedSumRateMaximization2024}, we propose a box-constrained movement mode for FAs. This approach ensures that constraints~\eqref{opt:tx_coupling_constraint} and~\eqref{opt:rx_coupling_constraint} are satisfied by maintaining a minimum gap $D$ between neighboring boxes. With this movement mode, constraints~\eqref{opt:tx_coupling_constraint} and~\eqref{opt:rx_coupling_constraint} are incorporated into~\eqref{opt:tx_antenna_constraint} and~\eqref{opt:rx_antenna_constraint}, respectively. Problems~\eqref{opt_tx_position:iteration} and~\eqref{opt_rx_position:problem} are reformulated as
\begin{equation}\label{opt_tx_position:problem_box}
 \underset{\mathbf{T}}{\max}\ \ f_{\rm Quad}\left(\mathbf{T}\right)\qquad {\text{s. t.}}\ \eqref{opt:tx_antenna_constraint}
\end{equation}
and
\begin{equation}\label{opt_rx_position:problem_box}          
 \underset{\mathbf{R}_k}{\max}\ \ f_{\rm Quad}\left(\mathbf{R}_k\right)\qquad {\text{s. t.}}\ \eqref{opt:rx_antenna_constraint},  
\end{equation}
respectively. In problems~\eqref{opt_tx_position:problem_box} and~\eqref{opt_rx_position:problem_box}, which adopt the box-constrained movement mode, the total number of inequality constraints increases linearly with the number of FAs. Specifically, problems~\eqref{opt_tx_position:problem_box} and~\eqref{opt_rx_position:problem_box} contain $M$ and $NK$ inequality constraints, respectively.

Since problems~\eqref{opt_tx_position:problem_box} and~\eqref{opt_rx_position:problem_box} have only cuboid boundaries as constraints, the optimal solutions are obtained by projecting the unconstrained optima $\mathbf{T}^\star$ and $\mathbf{R}_k^\star$ onto the cuboid regions~\cite{maMIMOCapacityCharacterization2024}. Thus, the closed-form solutions to problems~\eqref{opt_tx_position:problem_box} and~\eqref{opt_rx_position:problem_box} are given by
\begin{subequations}
    \begin{align}
 p_m^\mathtt{Tx} &= \min\Big(\max\left(p_m^{t, \star}, p_m^{\min}\right), p_m^{\max}\Big), \label{eq:opt_tx_position_proj}\\
 p_{kn}^\mathtt{Rx} &= \min\Big(\max\left(p_{kn}^{r, \star}, p_{kn}^{\min}\right), p_{kn}^{\max}\Big)\label{eq:opt_rx_position_proj},
    \end{align}
\end{subequations}
respectively, where $p$ denotes a spatial coordinate, and can be replaced by $x$, $y$, or $z$, depending on the dimension being optimized. Specifically, $p_m^{t, \star}$ and $p_{kn}^{r, \star}$ are the entries of the unconstrained optimal solutions $\mathbf{T}^\star$ and $\mathbf{R}_k^\star$, respectively. The box-constrained movement mode restricts the feasible domain of the problem~\eqref{opt:problem}. Although this approach sacrifices some achievable WSR for reduced complexity, we will show in Section~\ref{subsec:sim_box} that the degradation is negligible if the movable regions are sufficiently large. 

Based on the discussions above, we summarize the proposed BCA-based joint beamforming and antenna position optimization in Algorithm~\ref{alg:opt_overall}. In step~\ref{stp:init}, the beamforming matrices are initialized as $\mathbf{W}_k = \sqrt{\frac{P_{\rm max}}{Kd}} \left[\mathbf{I}_{d}, \mathbf{0}_{d\times (M-d)}\right]^\transpose$. Let $\rho$ denote the movable region of each antenna, normalized by $\lambda$, and the initial positions of transmit and receive FAs are uniform planar arrays with the spacing $\rho \lambda$. Notice that for the case of the decentralized implementation in Section~\ref{sec:dec_bca}, all the parameters are initialized similarly.

\begin{algorithm}
    \caption{Overall BCA-based algorithm for solving~\eqref{opt:problem_quadratic}}\label{alg:opt_overall}
    \begin{algorithmic}[1]
        \Require $M$, $N$, $K$, $P_{\max}$, $\alpha_k$, $\mathbf{\Sigma}_k$, $L_k^\mathtt{Tx}$, $L_k^\mathtt{Rx}$, $\theta_{ki}^\mathtt{Tx}$, $\phi_{ki}^\mathtt{Tx}$, $\theta_{kj}^\mathtt{Rx}$, $\phi_{kj}^\mathtt{Rx}$.
        \State Initialize $\underline{\mathbf{W}}$, $\mathbf{T}$, and $\underline{\mathbf{R}}$ to corresponding feasible values.\label{stp:init}
        \Repeat
        \State Update each $\mathbf{\Phi}_k$ via~\eqref{eq:opt_phi} and each $\mathbf{\Gamma}_k$ via~\eqref{eq:opt_gamma}.\label{stp:auxiliary}
        \State Update each $\mathbf{W}_k$ by bisection search via~\eqref{eq:opt_beamforming}.\label{stp:beamforming}
        \State Update $\mathbf{T}$ using MM according to Section~\ref{subsec:opt_tx_position}.
        \subState \qquad Calculate $\delta^{\mathtt{Tx}}$ via~\eqref{eq:opt_tx_position_delta}.
        \subState \qquad \textbf{repeat}
        \subState \qquad \qquad Calculate $\nabla_{\vect\left(\mathbf{T}\right)} f_{\rm Quad}\left(\mathbf{T}\right)$ via~\eqref{eq:opt_tx_position_dfdxyz}.
        \subState \qquad \qquad Calculate $\mathbf{T}^\star$ via~\eqref{eq:opt_tx_position_closed_form}.
        \subState \qquad \qquad Project $\mathbf{T}^\star$ onto cuboid regions via~\eqref{eq:opt_tx_position_proj}.
        \subState \qquad \textbf{until}{ the value of $f_{\rm Quad}\left(\mathbf{T}\right)$ converges.}
        
        \setcounter{subline}{0}
        \State Update $\mathbf{R}_k$ using MM according to Section~\ref{subsec:opt_rx_position}.
        \subState \qquad Calculate $\delta^{\mathtt{Rx}}_k$ via~\eqref{eq:opt_rx_position_delta}.
        \subState \qquad \textbf{repeat}
        \subState \qquad \qquad Calculate $\nabla_{\vect\left(\mathbf{R}_k\right)} f_{\rm Quad}\left(\mathbf{R}_k\right)$ via~\eqref{eq:opt_rx_position_dfdxyz}
        \subState \qquad \qquad Calculate $\mathbf{R}^\star_k$ via~\eqref{eq:opt_rx_position_closed_form}.
        \subState \qquad \qquad Project $\mathbf{R}_k^\star$ onto cuboid regions via~\eqref{eq:opt_rx_position_proj}.
        \subState \qquad \textbf{until}{ the value of $f_{\rm Quad}\left(\mathbf{R}_k\right)$ converges.}
        \Until{the value of $R$ converges.}
        \Ensure $\underline{\mathbf{W}}$, $\mathbf{T}$, $\underline{\mathbf{R}}$.
    \end{algorithmic}
\end{algorithm}

\subsection{Convergence Analysis}\label{subsec:convergence}
Since Algorithm~\ref{alg:opt_overall} is a two-loop algorithm, its convergence is ensured by the convergence of the MM-based inner loop and the BCA-based outer loop. To show the convergence of the proposed MM-based inner loop, we only discuss the convergence of the MM for transmit FA position optimization $\mathbf{T}$ for brevity. The convergence of the proposed MM algorithm is ensured by the monotonic increase and the upper boundedness of the objective function $f_{\rm Quad}(\mathbf{T})$. We introduce the following lemmas to demonstrate the convergence of the proposed MM algorithm.
\begin{lemma}\label{lemma:mono}
    Let $\overline{\mathbf{T}}$ and $\mathbf{T}$ be the FA position matrix before and after an MM iteration, respectively. Then, the objective value $f_{\rm Quad}$ increases monotonically, i.e.,
    \begin{equation}
 f_{\rm Quad}\left(\mathbf{T}\right) \geq f_{\rm Quad}\left(\overline{\mathbf{T}}\right),
    \end{equation}
\end{lemma}
\begin{IEEEproof}
    According to the discussions in Section~\ref{subsec:opt_tx_position}, the following inequalities hold:
    \begin{equation}
 f_{\rm Quad}\left(\mathbf{T}\right) \geq h^{\mathtt{Tx}}\left(\mathbf{T} \big\vert \overline{\mathbf{T}}\right) \geq h^{\mathtt{Tx}}\left(\overline{\mathbf{T}} \big\vert \overline{\mathbf{T}}\right) = f_{\rm Quad}\left(\overline{\mathbf{T}}\right),
    \end{equation}
 where the first inequality follows from the surrogate function property~\eqref{eq:mm_surrogate}, the second holds because $\mathbf{T}$ is the optimal solution to the problem~\eqref{opt_tx_position:problem_mm}, and the final equality results from the equality condition of the surrogate function at the expansion point $\overline{\mathbf{T}}$, given by~\eqref{eq:mm_equality}.
\end{IEEEproof}
\begin{lemma}\label{lemma:ub}
    The objective function $f_{\rm Quad}$ has finite upper bound as long as the constraints~\eqref{opt:power_constraint}--\eqref{opt:rx_coupling_constraint} hold, i.e., $\exists R_{\max}\in\mathbb{R}_+$, such that
    \begin{equation}
 f_{\rm Quad}(\underline{\mathbf{W}}, \mathbf{T}, \underline{\mathbf{R}}, \underline{\mathbf{\Gamma}}, \underline{\mathbf{\Phi}}) \leq R_{\max}
    \end{equation}
 for all $\underline{\mathbf{W}}, \mathbf{T}, \underline{\mathbf{R}}, \underline{\mathbf{\Gamma}}, \underline{\mathbf{\Phi}}$ satisfying~\eqref{opt:power_constraint}--\eqref{opt:rx_coupling_constraint}.
\end{lemma}
\begin{IEEEproof}
    Please refer to Appendix~\ref{appendix:upper_boundedness}.
\end{IEEEproof}
Given the above two lemmas, we conclude that the objective function $f_{\rm Quad}$ is monotonically increasing and upper bounded within each MM iteration. According to~\cite[Proposition~4]{shenOptimizationMIMODevicetoDevice2019}, the resulting solution $\mathbf{T}$ must be a stationary point, and the convergence of the proposed MM-based inner loop is thus guaranteed. The convergence of the BCA-based outer loop is a well-established result and is proved in~\cite[Section~V-D]{shenOptimizationMIMODevicetoDevice2019}. Therefore, the overall convergence of Algorithm~\ref{alg:opt_overall} is ensured.

\section{Decentralized Implementation of the BCA-Based Algorithm}\label{sec:dec_bca}
Although the proposed BCA-based algorithm in Section~\ref{sec:bca_sol} can effectively solve the problem~\eqref{opt:problem}, this centralized implementation suffers from high \emph{computational costs} as $M$ increases. The DBP architecture provides a promising solution to address the challenge by enabling DUs to solve small-scale subproblems in parallel. However, the advantages of the DBP architecture cannot be achieved without efficient decentralized optimization algorithms. To reduce the computational cost while maintaining similar performance, we propose a decentralized implementation of Algorithm~\ref{alg:opt_overall} based on the DBP architecture.
\subsection{DBP Architecture}
\begin{figure}[tb]
    \centering
    \includegraphics[width=0.47\textwidth]{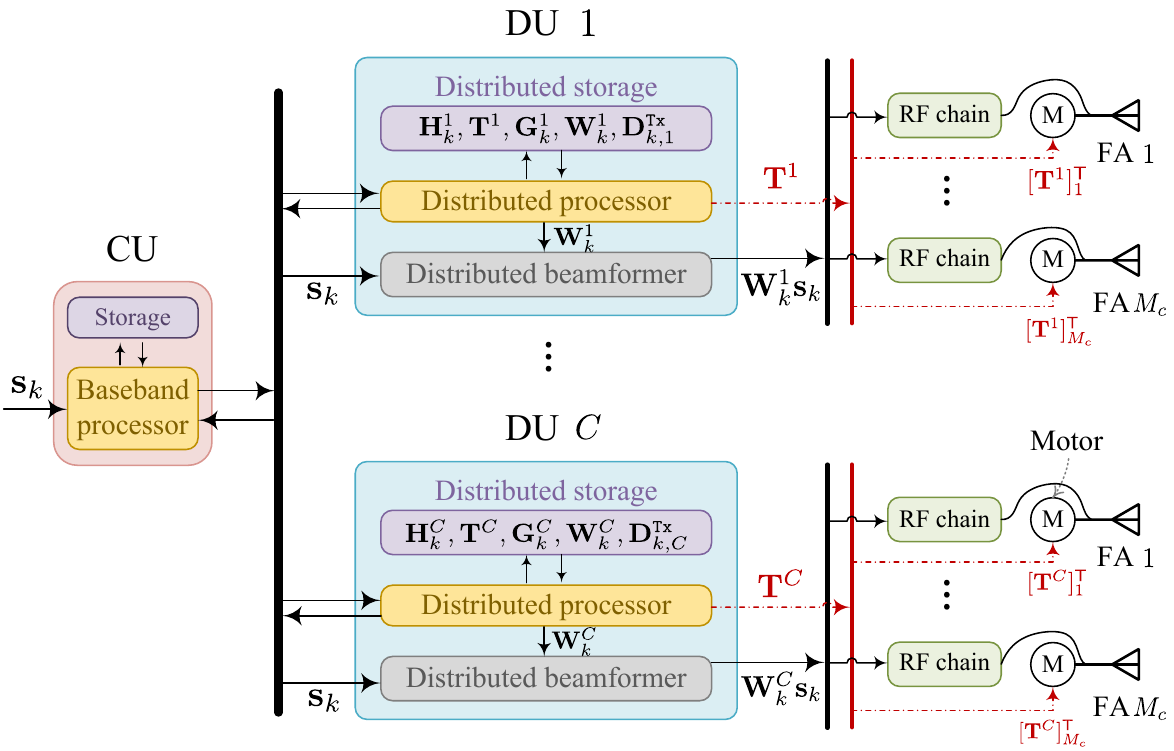}
    \caption{DBP architecture for FA-assisted MU-MIMO BS.}
    \label{fig:dec_arch}
    \vspace{-0.7\baselineskip}
\end{figure}
The DBP architecture for FA-assisted MU-MIMO system is illustrated in Fig.~\ref{fig:dec_arch}, where the black solid line represents the data signal, and the red dash-dotted line indicates the control signal. The DBP architecture is implemented on the BS side and partitions the transmit FA array into $C$ clusters. Each cluster contains $M_c$ transmit FAs with $M = C M_c$, and is managed by a DU equipped with dedicated RF circuitry, storage, and a baseband processor. With the DBP architecture, problem~\eqref{opt:problem} is solved cooperatively by the centralized unit (CU) and DUs, requiring data exchange between them. Specifically, the DUs compute the beamforming matrices $\mathbf{W}_k$ and transmit FA positions $\mathbf{T}$, enabling both beamforming and FA position control at the DU. To alleviate storage and interconnection costs, any matrix of dimension $M$ is stored distributively at the DUs. The letter $c$ denotes the submatrix stored across the $c$-th DU. The variables requiring distributive storage include the channel matrix $\mathbf{H}_k$, the positions of transmit FAs $\mathbf{T}$, the FRMs for transmit FAs $\mathbf{G}_k$, the beamforming matrices $\mathbf{W}_k$, and the transpose of the first-order derivative of $f_{\rm Quad}$ w.r.t. $\mathbf{G}_k$, denoted as $\mathbf{D}_k^{\mathtt{Tx}}$. Since each DU has access only to the positions of the transmit FAs it optimizes, FAs from different DUs may violate the constraint~\eqref{opt:tx_coupling_constraint}, potentially resulting in physical collisions. To address this issue, we adopt the box-constraint movement mode introduced in Section~\ref{subsec:bca_box}, which assigns independent and non-overlapping movable regions to each FA. To manage the \textit{computational cost}, we must ensure that the complexity at both the CU and DUs grows as a function of per-cluster antenna-count $M_c$, instead of $M$.\footnote{In practical systems, the number of DUs $C$ is typically fixed by the hardware design. The complexity at the CU and DUs is still proportional to $M$. However, as long as there is a sufficient number of DUs $C$, the complexity at the CU and DUs is no longer dominated by $M$.}

\begin{figure}[h]
    \centering
    \vspace{-0.2cm}
    \includegraphics[width=0.35\textwidth]{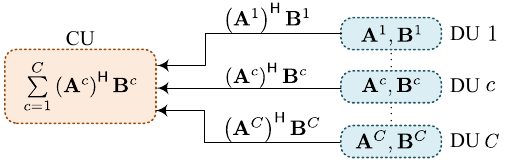}
    \caption{The decentralized calculation of $\sum_{c=1}^C (\mathbf{A}^c)^\hermconj \mathbf{B}^c$.}
    \vspace{-0.2cm}
    \label{fig:mul}
\end{figure}

Decentralized computation methods form the foundation of the proposed BCA-based decentralized algorithm. Among these methods, decentralized matrix multiplication is particularly critical and will be discussed in detail later. Given its pivotal role in our proposed algorithm, we first formalize this operation along with its implementation before presenting the algorithmic details. Suppose matrices $\mathbf{A}^c \in \mathbb{C}^{M_c \times a}$ and $\mathbf{B}^c \in \mathbb{C}^{M_c \times b}$ are stored at the $c$-th DU. We define the function $\mathsf{Mul}(\cdot, \cdot)$ as
\begin{equation}
 \mathsf{Mul}(\mathbf{A}^c, \mathbf{B}^c) = \sum\nolimits_{c=1}^C (\mathbf{A}^c)^\hermconj \mathbf{B}^c,
\end{equation}
whose result has dimensions independent of $M$ and can be efficiently stored and transmitted between the CU and DUs. The decentralized computation of $\mathsf{Mul}(\mathbf{A}^c, \mathbf{B}^c)$ is illustrated in Fig.~\ref{fig:mul}. Specifically, each DU first computes $(\mathbf{A}^c)^\hermconj \mathbf{B}^c$ locally and then transmits the result to the CU, where the final sum is aggregated.

\subsection{Decentralized Calculation of $\mathbf{M}_k$, $R$, and $f_{\rm Quad}$}\label{subsec:dec_mrf}

The values of $\mathbf{M}_k$, $R$, and $f_{\rm Quad}$, which are essential for evaluating system performance, are computed and stored at the CU. To compute them, we define $\tilde{\mathbf{G}}_{kj} \triangleq \mathsf{Mul}\Big(\big(\mathbf{G}^c_k(\overline{\mathbf{T}}^c)\big)^\hermconj, \overline{\mathbf{W}}_j^c\Big)$. Once the CU obtains $\tilde{\mathbf{G}}_{kj}$ as illustrated in Fig.~\ref{fig:mul}, the values of $\mathbf{M}_k$, $R$, and $f_{\rm Quad}$ are calculated directly at the CU:
\begin{equation}
 \mathbf{M}_k = \sum_{j=1, j\neq k}^K\overline{\mathbf{F}}_k^\hermconj\mathbf{\Sigma}_k\tilde{\mathbf{G}}_{kk}\tilde{\mathbf{G}}_{kk}^\hermconj\mathbf{\Sigma}_k^\hermconj\overline{\mathbf{F}}_k + \sigma_k^2\mathbf{I},
\end{equation}
\begin{equation}
 R = \sum_{k=1}^K \alpha_k \log\det\left(\mathbf{I} + \overline{\mathbf{F}}_k^\hermconj\mathbf{\Sigma}_k\tilde{\mathbf{G}}_{kk}\mathbf{M}_k^{-1}\tilde{\mathbf{G}}_{kk}^\hermconj\mathbf{\Sigma}_k^\hermconj\overline{\mathbf{F}}_k + \sigma_k^2\mathbf{I}\right),
\end{equation}
and~\eqref{eq:dec_quadratic_trans} at the bottom of the page, respectively.
\begin{figure*}[hb]
    \vspace{-\baselineskip}
    \hrulefill
    \normalsize
    \setcounter{MYtempeqncnt}{\value{equation}}
    \setcounter{equation}{\value{MYtempeqncnt}}
    \begin{multline}\label{eq:dec_quadratic_trans}
 f_{\rm Quad}\left(\underline{\mathbf{W}}, \mathbf{T}, \underline{\mathbf{R}}, \underline{\mathbf{\Gamma}}, \underline{\mathbf{\Phi}}\right) =
        \sum_{k=1}^{K}\left(\alpha_k\log\det\left(\mathbf{I}+\mathbf{\Gamma}_k\right)-\alpha_k\trace\left(\mathbf{\Gamma}_k\right) \right.\\
 + \trace\Big[\left(\mathbf{I}+\mathbf{\Gamma}_k\right)\Big(\sqrt{\alpha_k}\tilde{\mathbf{G}}_{kk}^\hermconj\mathbf{\Sigma}_k^\hermconj\mathbf{F}_k\mathbf{\Phi}_k + \sqrt{\alpha_k}\mathbf{\Phi}_k^{\hermconj}\mathbf{F}_k^\hermconj\mathbf{\Sigma}_k\tilde{\mathbf{G}}_{kk} -\mathbf{\Phi}_k^{\hermconj}\Big(\sum_{j=1}^{K}\overline{\mathbf{F}}_k^\hermconj\mathbf{\Sigma}_k\tilde{\mathbf{G}}_{kk}\tilde{\mathbf{G}}_{kk}^\hermconj\mathbf{\Sigma}_k^\hermconj\overline{\mathbf{F}}_k+\sigma_k^2\mathbf{I}\Big)\mathbf{\Phi}_k\Big)\Big]\Big).
    \end{multline}
    \setcounter{equation}{\value{equation}}
\end{figure*}
\subsection{Decentralized Update of $\underline{\mathbf{\Gamma}}$ and $\underline{\mathbf{\Phi}}$}\label{subsec:dec_auxiliary}
The values of $\underline{\mathbf{\Gamma}}$ and $\underline{\mathbf{\Phi}}$ are computed and stored at the CU. Additionally, they can be transmitted to and stored at the DUs. With $\tilde{\mathbf{G}}_{kj}$, the remaining computations for updating $\underline{\mathbf{\Gamma}}$ and $\underline{\mathbf{\Phi}}$ can be performed directly at the CU:
\begin{align}
 \mathbf{\Gamma}_k &= \tilde{\mathbf{G}}_{kk}^\hermconj\mathbf{\Sigma}_k^\hermconj\overline{\mathbf{F}}_k \overline{\mathbf{M}}_k^{-1}\overline{\mathbf{F}}_k^\hermconj\mathbf{\Sigma}_k\tilde{\mathbf{G}}_{kk}, \label{eq:dec_gamma}\\
 \mathbf{\Phi}_k &= \sqrt{\alpha_k}\left(\overline{\mathbf{M}}_k+\overline{\mathbf{F}}_k^\hermconj\mathbf{\Sigma}_k\tilde{\mathbf{G}}_{kk}\tilde{\mathbf{G}}_{kk}^\hermconj\mathbf{\Sigma}_k^\hermconj\overline{\mathbf{F}}_k\right)^{-1}\overline{\mathbf{F}}_k^\hermconj\mathbf{\Sigma}_k\tilde{\mathbf{G}}_{kk}.\label{eq:dec_phi}
\end{align}

\subsection{Decentralized Update of $\underline{\mathbf{W}}$}\label{subsec:dec_w}

\begin{figure*}[ht]
    \normalsize
    \setcounter{MYtempeqncnt}{\value{equation}}
    \setcounter{equation}{60}
    \begin{multline}\label{eq:nonh_trans}
 f_{\rm NonH}\left(\underline{\mathbf{W}}, \underline{\mathbf{\Psi}}\right) = \sum_{k=1}^K \trace\Big[-\eta\mathbf{W}_k^\hermconj\mathbf{W}_k - 2\Re\Big(\mathbf{W}_k^\hermconj\Big(\sum_{j=1}^K{\mathbf{H}_j^\hermconj\mathbf{\Phi}_j\left(\mathbf{I}+\mathbf{\Gamma}_j\right)\mathbf{\Phi}_j^\hermconj\mathbf{H}_j} - \eta\mathbf{I}\Big)\mathbf{\Psi}_k\Big) \\
 - \mathbf{\Psi}_k^\hermconj\Big(\eta\mathbf{I} - \sum_{j=1}^K{\mathbf{H}_j^\hermconj\mathbf{\Phi}_j\left(\mathbf{I}+\mathbf{\Gamma}_j\right)\mathbf{\Phi}_j^\hermconj\mathbf{H}_j}\Big)\mathbf{\Psi}_k + \left(\mathbf{I} + \mathbf{\Gamma}_k\right)\left(\sqrt{\alpha_k}\mathbf{W}_k^\hermconj \mathbf{H}_k^\hermconj \mathbf{\Phi}_k + \sqrt{\alpha_k}\mathbf{\Phi}_k\mathbf{H}_k\mathbf{W}_k\right)\Big] + \mathrm{const}.
    \end{multline}
    \vspace{-0.5\baselineskip}
    \hrulefill
    \setcounter{equation}{\value{MYtempeqncnt}}
\end{figure*}

A major challenge in the decentralized update of $\underline{\mathbf{W}}$ is the matrix inversion in~\eqref{eq:opt_beamforming}. Since matrix inversion of dimension $M$ cannot be computed distributively, it is necessary to avoid matrix inversion in~\eqref{eq:opt_beamforming} to achieve a decentralized update of $\underline{\mathbf{W}}$. Therefore, we first discuss an inverse-free update step of $\underline{\mathbf{W}}$, followed by the proposed decentralized implementation of the algorithm.

If the quadratic term's coefficient matrix w.r.t. $\mathbf{W}_k$ is diagonal, its inversion reduces to element-wise reciprocation of the diagonal entries, which can be computed distributively. Therefore, we introduce the following proposition to transform the coefficient matrix w.r.t. $\mathbf{W}_k$ into a diagonal form.

\begin{proposition}[Matrix non-homogeneous transform~{\cite[Corollary~19]{zhangDiscerningEnhancingWeighted2023}}]\label{prop:nonhomogeneous_transform}
 For Hermitian matrices $\mathbf{L}$ and $\mathbf{M}$ satisfying $\mathbf{M} \succeq \mathbf{L}$, the problem
    \begin{equation}\label{eq:nonh_obj}
 \underset{\mathbf{X}\in\mathcal{X}}{\max}\ -\trace\left(\mathbf{X}^\hermconj\mathbf{L}\mathbf{X}\right)
    \end{equation}
 is equivalent to
    \begin{footnotesize}
        \begin{equation}\label{eq:nonh_obj_trans}
 \underset{\mathbf{X}, \mathbf{\Psi}\in\mathcal{X}}{\max}\ -\trace\left(\mathbf{X}^\hermconj\mathbf{M}\mathbf{X} + 2\Re\left(\mathbf{X}^\hermconj \left(\mathbf{L}-\mathbf{M}\right)\mathbf{\Psi}\right) + \mathbf{\Psi}^\hermconj\left(\mathbf{M}-\mathbf{L}\right)\mathbf{\Psi}\right),
        \end{equation}
    \end{footnotesize}%
 in the sense that they achieve identical optimal objective values with identical optimal solutions, where $\mathbf{\Psi}$ is introduced as an auxiliary variable. 
\end{proposition}

By applying Proposition~\ref{prop:nonhomogeneous_transform} to $f_{\rm Quad}\left(\underline{\mathbf{W}}\right)$ and setting $\mathbf{M} = \eta\mathbf{I}$, where
\begin{equation}\label{eq:nonh_eta}
    \eta = \Big\lVert\sum\nolimits_{j=1}^K{\overline{\mathbf{H}}_j^\hermconj\overline{\mathbf{\Phi}}_j\left(\mathbf{I}+\overline{\mathbf{\Gamma}}_j\right)\overline{\mathbf{\Phi}}_j^\hermconj\overline{\mathbf{H}}_j}\Big\rVert_{\rm F},
\end{equation}
the problem~\eqref{opt_beamforming:problem} is reformulated as~\cite{zhangDiscerningEnhancingWeighted2023}%\vspace{0.2cm}
\begin{equation}\label{opt:problem_nonh}
 \underset{\underline{\mathbf{W}}, \underline{\mathbf{\Psi}}}{\max}\ \ f_{\rm NonH}\left(\underline{\mathbf{W}}, \underline{\mathbf{\Psi}}\right)\qquad {\text{s. t.}}\ \eqref{opt:power_constraint},
\end{equation}
where $f_{\rm NonH}\left(\underline{\mathbf{W}}, \underline{\mathbf{\Psi}}\right)$ is given as~\eqref{eq:nonh_trans} at the top of the next page. For simplicity, we define $\underline{\mathbf{\Psi}} = \left\{\mathbf{\Psi}_k, \forall k\right\}$.
\addtocounter{equation}{1}
Since the problem~\eqref{opt:problem_nonh} involves both the auxiliary variables $\underline{\mathbf{\Psi}}$ and the beamforming matrices $\underline{\mathbf{W}}$, we update one set of variables while keeping the other fixed.

First, we update $\underline{\mathbf{\Psi}}$ while keeping $\underline{\mathbf{W}}$ fixed. According to Proposition~\ref{prop:nonhomogeneous_transform}, the optimal $\mathbf{\Psi}_k$ is given by
\begin{equation}\label{eq:opt_beamforming_nonh_psi}
 \mathbf{\Psi}_k = \overline{\mathbf{W}}_k.
\end{equation}
Next, we update $\underline{\mathbf{W}}$ while keeping $\underline{\mathbf{\Psi}}$ fixed. By substituting~\eqref{eq:opt_beamforming_nonh_psi} into~\eqref{eq:nonh_trans} and setting the first-order derivative of $f_{\rm NonH}\left(\underline{\mathbf{W}}, \overline{\underline{\mathbf{W}}}\right)$ w.r.t. $\mathbf{W}_k$ to zero, we obtain the closed-form solution to the problem~\eqref{opt:problem_nonh}:
\begin{equation}\label{eq:opt_beamforming_w_nonh}
 \mathbf{W}_k = \mathbf{Q}_k\min\Big\{\sqrt{P_{\max}/P_Q}, 1\Big\},
\end{equation}
where $P_Q$ denotes the transmission power with beamforming matrix $\mathbf{Q}_k$:
\begin{equation}
 P_Q = \sum\nolimits_{j=1}^K\trace\left(\mathbf{Q}_j^\hermconj \mathbf{Q}_j\right),
\end{equation}
and the matrix $\mathbf{Q}_k$ denotes the beamforming matrix without the power constraint:
\begin{align}\label{eq:opt_beamforming_nonh_q}
 \mathbf{Q}_k =& \eta^{-1}\Big[\sqrt{\alpha_k}\overline{\mathbf{H}}_k^\hermconj\overline{\mathbf{\Phi}}_k\left(\mathbf{I} + \overline{\mathbf{\Gamma}}_k\right) \nonumber\\
    &- \Big(\sum_{j=1}^K\overline{\mathbf{H}}_j^\hermconj \overline{\mathbf{\Phi}}_j \left(\mathbf{I}+\overline{\mathbf{\Gamma}}_j\right) \overline{\mathbf{\Phi}}_j^\hermconj \overline{\mathbf{H}}_j - \eta \mathbf{I}\Big)\overline{\mathbf{W}}_k\Big].
\end{align}

Although the per-iteration complexity is significantly reduced by eliminating matrix inversions, more iterations are required for convergence, which may still be time-consuming. Therefore, it is necessary to reduce the number of iterations. As demonstrated in~\cite{shenAcceleratingQuadraticTransform2024}, we can reduce the number of iterations by using momentum methods. Among these methods, Nesterov's extrapolation strategy~\cite{nesterovLecturesConvexOptimization2018} extrapolates $\mathbf{W}_k$ along the direction defined by the two previous iterations, $\overline{\mathbf{W}}_k$ and $\overline{\overline{\mathbf{W}}}_k$, to predict its value in the following iteration. This approach is effective in this scenario~\cite{shenAcceleratingQuadraticTransform2024}. The extrapolated value $\mathbf{\Upsilon}_k$ is defined as
\begin{equation}\label{eq:Upsilon}
 \mathbf{\Upsilon}_k \triangleq \overline{\mathbf{W}}_k + \nu_i\left(\overline{\mathbf{W}}_k - \overline{\overline{\mathbf{W}}}_k\right),
\end{equation}
where $\nu_i = \max\left\{(i-2)/(i+1), 0\right\}$ represents the extrapolation step size in the $i$-th BCA iteration. 
Using Nesterov's extrapolation strategy, the matrix $\mathbf{Q}_k$ is calculated as
\begin{align}\label{eq:opt_beamforming_extr_q}
 \mathbf{Q}_k =& \eta^{-1}\Big[\sqrt{\alpha_k}\overline{\mathbf{H}}_k^\hermconj\overline{\mathbf{\Phi}}_k\left(\mathbf{I} + \overline{\mathbf{\Gamma}}_k\right) \nonumber\\
    &- \Big(\sum_{j=1}^K\overline{\mathbf{H}}_j^\hermconj \overline{\mathbf{\Phi}}_j \left(\mathbf{I}+\overline{\mathbf{\Gamma}}_j\right) \overline{\mathbf{\Phi}}_j^\hermconj \overline{\mathbf{H}}_j - \eta \mathbf{I}\Big)\mathbf{\Upsilon}_k\Big].
\end{align}
Then, the matrix $\mathbf{W}_k$ is obtained by substituting~\eqref{eq:opt_beamforming_extr_q} into~\eqref{eq:opt_beamforming_w_nonh}.
\addtocounter{equation}{2}

After obtaining the improved closed-form expression of $\underline{\mathbf{W}}$ given in~\eqref{eq:opt_beamforming_w_nonh} and~\eqref{eq:opt_beamforming_extr_q}, the decentralized update of $\underline{\mathbf{W}}$ can be achieved correspondingly. First, we compute $\eta$ in a distributed manner. Since $\mathbf{I} + \overline{\mathbf{\Gamma}}_k \succ \mathbf{0}$, we perform the eigenvalue decomposition (EVD) to obtain $\mathbf{\Lambda}_k \in \mathbb{R}_{+}^{d \times d}$, a diagonal matrix of eigenvalues, and $\mathbf{\Xi}_k \in \mathbb{C}^{d \times d}$, whose columns are the corresponding eigenvectors. The CU broadcasts $\mathbf{\Lambda}_k$ and $\mathbf{\Xi}_k$ to all DUs, and each DU computes $\mathbf{P}_k^c$ as
\begin{equation}
    \mathbf{P}_k^c \triangleq (\overline{\mathbf{H}}_k^c)^\hermconj \overline{\mathbf{\Phi}}_k \mathbf{\Xi}_k \sqrt{\mathbf{\Lambda}_k}.
\end{equation}
Then, the value $\tilde{\mathbf{P}}_{kj} \triangleq \mathsf{Mul}(\mathbf{P}_k^c, \mathbf{P}_j^c)$ is calculated similar to the process in Fig.~\ref{fig:mul}, and the CU computes $\eta$ as
\begin{equation}\label{eq:eta_calc}
    \eta = \sqrt{\sum\nolimits_{j=1}^K \sum\nolimits_{k=1}^K \trace\left(\tilde{\mathbf{P}}_{kj}^\hermconj \tilde{\mathbf{P}}_{kj}\right)}.
\end{equation}
Note that~\eqref{eq:eta_calc} is equivalent to~\eqref{eq:nonh_eta} and is derived using the trace property of the Frobenius norm.

\begin{figure}[H]
    \vspace{-0.2cm}
    \centering
    \includegraphics[width=0.45\textwidth]{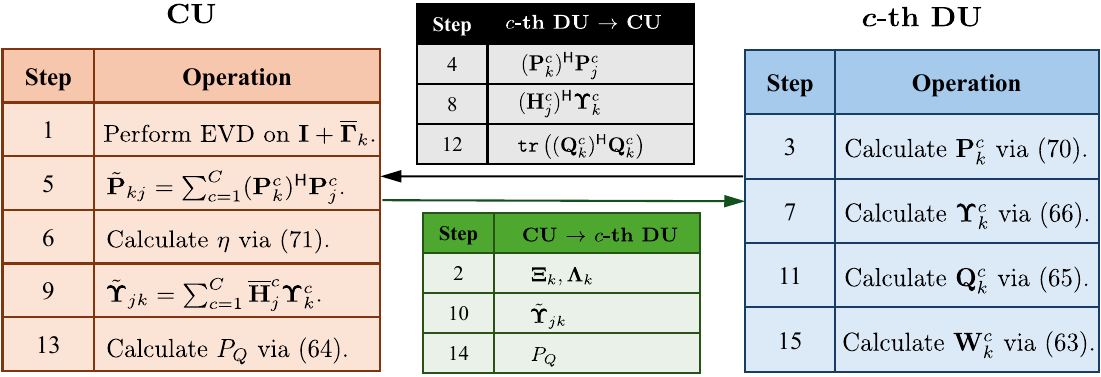}
    \caption{The decentralized calculation of $\underline{\mathbf{W}}$ under the DBP architecture.}
    \label{fig:bf}
\end{figure}

Next, we use the previously computed $\eta$ to calculate $\mathbf{W}_k^c$. To avoid repetition of formulas, we refer directly to the centralized equations introduced earlier. The only modification in the decentralized setting is to append a superscript $c$ to the terms $\overline{\mathbf{H}}_k$, $\overline{\mathbf{W}}_k$, $\overline{\overline{\mathbf{W}}}_k$, $\mathbf{\Upsilon}_k$, and $\mathbf{Q}_k$. First, we compute the extrapolated beamforming matrices $\mathbf{\Upsilon}_k^c$ at the $c$-th DU by~\eqref{eq:Upsilon}. Then, we compute $\tilde{\mathbf{\Upsilon}}_{jk}$ similar to Fig.~\ref{fig:mul} and compute $\mathbf{Q}_k^c$ via~\eqref{eq:opt_beamforming_nonh_q}. To calculate the value of $P_Q$, each DU locally computes $\trace(\left(\mathbf{Q}_k^c\right)^\hermconj \mathbf{Q}_k^c)$ and transmits the result to the CU. The CU then aggregates the received values by $P_Q = \sum\nolimits_{k=1}^K\sum\nolimits_{c=1}^C \trace\left(\left(\mathbf{Q}_k^c\right)^\hermconj \mathbf{Q}_k^c\right)$. Once $P_Q$ is obtained, it is broadcast to all DUs, and $\mathbf{W}^c_k$ can be computed at the $c$-th DU by~\eqref{eq:opt_beamforming_w_nonh}. The aforementioned process is summarized in Fig.~\ref{fig:bf}, where only the $c$-th DU is shown for brevity.

\subsection{Decentralized Update of $\mathbf{T}$ and $\underline{\mathbf{R}}$}\label{subsec:dec_pos}
\begin{figure*}[ht]
    \normalsize
    \setcounter{MYtempeqncnt}{\value{equation}}
    \setcounter{equation}{74}
    \begin{small}
    \begin{equation}\label{eq:dec_opt_tx_position_delta}
        \begin{aligned}
            \delta^{\mathtt{Tx}}_{m, c} = \frac{24\pi^2}{\lambda^2}&\sum_{k=1}^K L_k^\mathtt{Tx}\Big[\Big(\sum_{t=1}^K\Big\lVert[\overline{\mathbf{W}}_t^c]_m\Big\rVert_2 \sum_{j=1}^M \Big\lVert[\overline{\mathbf{W}}_t]_j\Big\rVert_2 + \sum_{t=1}^K\sum_{s=1}^K[\overline{\mathbf{W}}_t^c]_m\tilde{\mathbf{W}}_{ts}[\overline{\mathbf{W}}_s^c]_m^\hermconj\Big)\lVert\hat{\mathbf{\Sigma}}_k^{\mathtt{Tx}}\rVert_2 + \sqrt{\frac{\alpha_k}{L_k^\mathtt{Tx}}} \Big\lVert[\overline{\mathbf{W}}_k^c]_m\left(\mathbf{I}+\overline{\mathbf{\Gamma}}_k\right)\overline{\mathbf{\Phi}}_k^\hermconj\overline{\mathbf{F}}_k^\hermconj\mathbf{\Sigma}_k^\hermconj\Big\rVert_2\Big].
        \end{aligned}    
    \end{equation}
    \end{small}%
    \hrulefill
    \vspace{-0.3cm}
    \setcounter{equation}{\value{MYtempeqncnt}}
\end{figure*}

\begin{figure*}[ht]
    \normalsize
    \setcounter{MYtempeqncnt}{\value{equation}}
    \setcounter{equation}{75}
    \begin{equation}\label{eq:dec_opt_rx_position_delta}
        \begin{aligned}
            \delta^{\mathtt{Rx}}_k = \underset{1\leq n\leq N}{\max}\frac{24\pi^2}{\lambda^2} &L_k^\mathtt{Rx}\left(\left(\sum\nolimits_{j=1}^N\Big\lvert[\overline{\mathbf{\Phi}}_k]_n\left(\mathbf{I} + \overline{\mathbf{\Gamma}}_k\right)[\overline{\mathbf{\Phi}}_k]_j^\hermconj\Big\rvert + \sqrt{N}\Big\lVert[\overline{\mathbf{\Phi}}_k]_n\left(\mathbf{I} + \overline{\mathbf{\Gamma}}_k\right)\overline{\mathbf{\Phi}}_k^\hermconj\Big\rVert_2\right)\Big\lVert\sum\nolimits_{t=1}^K \mathbf{\Sigma}_k \tilde{\mathbf{G}}_{kt}\tilde{\mathbf{G}}_{kt}^\hermconj \mathbf{\Sigma}_k^\hermconj\Big\rVert_2\right.\\
            &\left. + \sqrt{\frac{\alpha_k}{L_k^\mathtt{Rx}}}\Big\lVert[\mathbf{\Phi}_k]_n\left(\mathbf{I} + \overline{\mathbf{\Gamma}}_k\right)\tilde{\mathbf{G}}_{kk}^\hermconj\mathbf{\Sigma}_k^\hermconj\Big\rVert_2\right).
        \end{aligned}    
    \end{equation}
    \hrulefill
    \vspace{-0.3cm}
    \setcounter{equation}{\value{MYtempeqncnt}}
\end{figure*}
To update $\mathbf{T}$ and $\underline{\mathbf{R}}$ distributively, we must compute $\mathbf{D}_k^\mathtt{Tx}$, $\mathbf{D}_k^\mathtt{Rx}$, $\delta^{\mathtt{Tx}}$, and $\delta^{\mathtt{Rx}}_k$ distributively in advance. The matrix $\mathbf{D}_k^\mathtt{Tx}$ is stored distributively as $\mathbf{D}_{k, c}^\mathtt{Tx}$, whereas $\mathbf{D}_k^\mathtt{Rx}$ is stored at the CU. To calculate $\mathbf{D}_k^\mathtt{Tx}$, the CU broadcasts $\overline{\mathbf{\Gamma}}_k$, $\overline{\mathbf{\Phi}}_k$, $\overline{\mathbf{F}}_k$, and $\tilde{\mathbf{G}}_{kj}$ to each DU. Then, the $c$-th DU computes $\mathbf{D}_{k, c}^\mathtt{Tx}$ as
\begin{align}\label{eq:dec_opt_tx_position_D}
 \mathbf{D}_{k, c}^\mathtt{Tx} =& \sqrt{\alpha_k}\overline{\mathbf{W}}_k^c\left(\mathbf{I} + \overline{\mathbf{\Gamma}}_k\right)\overline{\mathbf{F}}_k^\hermconj\mathbf{\Sigma}_k \nonumber \\
    &- \sum\nolimits_{j=1}^K\overline{\mathbf{W}}_j^c\tilde{\mathbf{G}}_{kj}^\hermconj\mathbf{\Sigma}_k^\hermconj\overline{\mathbf{F}}_k\overline{\mathbf{\Phi}}_k\left(\mathbf{I} + \overline{\mathbf{\Gamma}}_k\right)\overline{\mathbf{\Phi}}_k^\hermconj\overline{\mathbf{F}}_k^\hermconj\mathbf{\Sigma}_k.
\end{align}
Meanwhile, as indicated by~\eqref{eq:opt_rx_position_D}, calculating $\mathbf{D}_k^\mathtt{Rx}$ requires only $\tilde{\mathbf{G}}_{kj}$. The remaining calculations in~\eqref{eq:opt_rx_position_D} are performed at the CU, given by
\begin{align}\label{eq:dec_opt_rx_position_D}
 \mathbf{D}^{\mathtt{Rx}}_k =& \sqrt{\alpha_k}\overline{\mathbf{\Phi}}_k\left(\mathbf{I} + \overline{\mathbf{\Gamma}}_k\right)\tilde{\mathbf{G}}_{kk}^\hermconj \mathbf{\Sigma}_k^\hermconj \nonumber\\
    &- \overline{\mathbf{\Phi}}_k \left(\mathbf{I} + \overline{\mathbf{\Gamma}}_k\right) \overline{\mathbf{\Phi}}_k^\hermconj \overline{\mathbf{F}}_k^\hermconj\mathbf{\Sigma}_k \sum\nolimits_{j=1}^K \tilde{\mathbf{G}}_{kj}\tilde{\mathbf{G}}_{kj}^\hermconj \mathbf{\Sigma}_k^\hermconj.
\end{align}

Scalars $\delta^{\mathtt{Tx}}$ and $\delta_k^\mathtt{Rx}$ are stored at the CU. To compute the values distributively, we define $\tilde{\mathbf{W}}_{kj} \triangleq \mathsf{Mul}(\overline{\mathbf{W}}_k^c, \overline{\mathbf{W}}_j^c)$ and it is calculated as shown in Fig.~\ref{fig:mul}. However, $\delta^{\mathtt{Tx}}$ still cannot be computed distributively since $\hat{\mathbf{W}}$ in~\eqref{eq:opt_tx_position_delta} cannot be stored and calculated. We thus apply the triangle inequality and Cauchy-Schwarz inequality to $\sum_{j=1}^M \big\lvert[\hat{\mathbf{W}}]_{mj}\big\rvert$ and obtain
\begin{equation}\label{eq:dec_opt_tx_position_w_norm}
    \sum_{j=1}^M \Big\lvert[\hat{\mathbf{W}}]_{mj}\Big\rvert \leq \sum_{k=1}^K\Big\lVert[\overline{\mathbf{W}}_k^c]_m\Big\rVert_2 \sum_{c^\prime=1}^C\sum_{j=1}^{M_c} \Big\lVert[\overline{\mathbf{W}}_k^{c^\prime}]_j\Big\rVert_2.
\end{equation}
To calculate the r.h.s. of~\eqref{eq:dec_opt_tx_position_w_norm}, each DU first computes $\sum_{j=1}^{M_c} \big\lVert[\overline{\mathbf{W}}_k^c]_j\big\rVert_2$ and transmits it to the CU. Then, the CU calculates the summation $\sum_{c=1}^C\sum_{j=1}^{M_c} \big\lVert[\overline{\mathbf{W}}_k^c]_j\big\rVert_2$ and broadcasts it back to DUs, where the r.h.s. of~\eqref{eq:dec_opt_tx_position_w_norm} is finally computed. Plugging~\eqref{eq:dec_opt_tx_position_w_norm} into~\eqref{eq:opt_tx_position_delta}, we obtain the expression for $\delta^{\mathtt{Tx}}$. To better demonstrate the decentralized computation of $\delta^{\mathtt{Tx}}$, we define $\delta^{\mathtt{Tx}} = \underset{1\leq c\leq C}{\max}\underset{1\leq m\leq M_c}{\max} \delta^{\mathtt{Tx}}_{m, c}$, and the expression of $\delta^{\mathtt{Tx}}_{m, c}$ is given by~\eqref{eq:dec_opt_tx_position_delta} at the top of the next page. With $\tilde{G}_{kj}$ computed previously, the value of $\delta^{\mathtt{Rx}}_k$ can be directly calculated at the CU, as shown in~\eqref{eq:dec_opt_rx_position_delta} at the top of the next page.

\addtocounter{equation}{2}

\begin{figure}[H]
    \centering
    \includegraphics[width=0.45\textwidth]{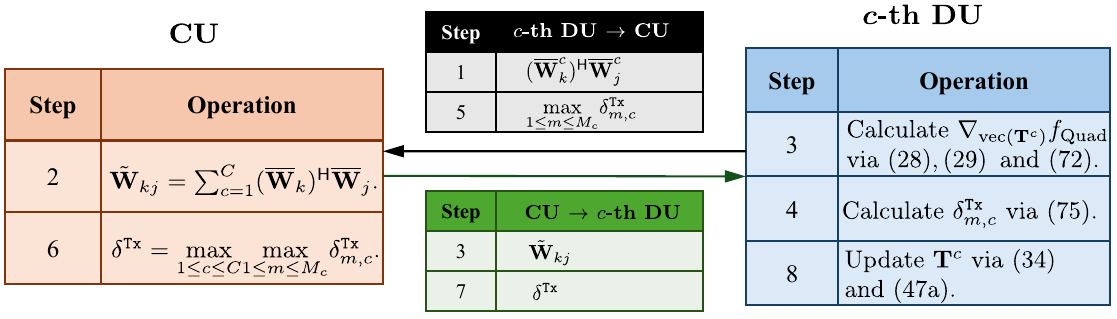}
    \caption{The decentralized update of $\mathbf{T}$ under the DBP architecture.}
    \label{fig:tx_pos}
\end{figure}

Here, we summarize the update of $\underline{\mathbf{T}}$ under the DBP architecture. We refer directly to the centralized equations introduced earlier to avoid presenting similar formulas. The only modification in the decentralized setting is to append a superscript $c$ to the variable $\mathbf{t}_m$ and a subscript $c$ to $\mathbf{D}_k^{\mathtt{Tx}}$. First, we compute the coefficients of the \textit{surrogate function} $h^{\mathtt{Tx}}(\mathbf{T}\big\vert\underline{\overline{\mathbf{T}}})$. This begins with the distributed computation of $\tilde{\mathbf{W}}_{kj}$ as illustrated by Fig.~\ref{fig:mul}. After the CU broadcasts $\tilde{\mathbf{W}}_{kj}$ to all DUs, the entries of $\nabla_{\vect(\mathbf{T}^c)}f_{\rm Quad}(\overline{\mathbf{T}}^c)$ are calculated by~\eqref{eq:opt_tx_position_dfdxyz},~\eqref{eq:opt_tx_position_xi}, and~\eqref{eq:dec_opt_tx_position_D} at each DU. Next, to compute $\delta^\mathtt{Tx}$, each DU evaluates $\delta^{\mathtt{Tx}}_{m, c}$ by~\eqref{eq:dec_opt_tx_position_delta} and selects the greatest value to send to the CU, which determines the largest and sets it as $\delta^{\mathtt{Tx}}$. With these coefficients computed, the optimal solution for $\underline{\mathbf{T}}$ is computed at the DUs using~\eqref{eq:opt_tx_position_closed_form} and~\eqref{eq:opt_tx_position_proj}. The aforementioned process is summarized in Fig.~\ref{fig:tx_pos}, where only the $c$-th DU is shown for brevity.

The calculation of $\mathbf{R}_{kn}$ can be directly executed at the CU. First, we compute the coefficients of the \textit{surrogate function} $h^{\mathtt{Rx}}_k(\mathbf{R}_k\big\vert\overline{\mathbf{R}}_k)$. Leveraging the $\tilde{\mathbf{G}}_{kj}$ computed previously, the CU computes the entries of $\nabla_{\vect(\mathbf{R}_k)} f_{\rm Quad}(\overline{\mathbf{R}}_k)$ and $\delta_k^\mathtt{Rx}$ using~\eqref{eq:opt_rx_position_dfdxyz} and~\eqref{eq:dec_opt_rx_position_delta}, respectively. Then, the optimal solution to problem~\eqref{opt_rx_position:problem_box} is computed at the CU via~\eqref{eq:opt_rx_position_closed_form} and~\eqref{eq:opt_rx_position_proj}. The key steps of the above decentralized DBP-based algorithm are summarized in Algorithm~\ref{alg:dec_opt_overall}.
\begin{algorithm}
    \caption{Decentralized Implementation of Algorithm~\ref{alg:opt_overall}}\label{alg:dec_opt_overall}
    \begin{algorithmic}[1]
        \Require {\small $C$, $M$, $N$, $K$, $P_{\max}$, $\alpha_k$, $\mathbf{\Sigma}_k$, $L_k^\mathtt{Tx}$, $L_k^\mathtt{Rx}$, $\theta_{ki}^\mathtt{Tx}$, $\phi_{ki}^\mathtt{Tx}$, $\theta_{kj}^\mathtt{Rx}$, $\phi_{kj}^\mathtt{Rx}$.}
        \State Initialize $\underline{\mathbf{W}}$, $\mathbf{T}$, and $\underline{\mathbf{R}}$ to corresponding feasible values.
        \Repeat
        \State Update each $\mathbf{\Phi}_k$ and $\mathbf{\Gamma}_k$ distributively according to the
        \Statex \hspace{1.1em} steps in Section~\ref{subsec:dec_auxiliary} and store the results at the CU.
        \State Update each \hspace{0.2em} $\mathbf{W}_k^c$ \hspace{0.1em} distributively \hspace{0.1em} according \hspace{0.05em} to \hspace{0.05em} the 
        \Statex \hspace{1.1em} steps in Section~\ref{subsec:dec_w} and store it at the $c$-th DU.
        \State Update \hspace{0.2em}$\mathbf{T}^c$\hspace{0.2em} and \hspace{0.1em} $\mathbf{R}_k$ \hspace{0.1em} distributively according to the 
        \Statex \hspace{1.1em} steps in Section~\ref{subsec:dec_pos}, \hspace{0.1em} and store them at the $c$-th DU 
        \Statex \hspace{1.1em} and the CU, respectively.
        \Until{the value of $R$ converges.}
        \Ensure $\underline{\mathbf{W}}$, $\mathbf{T}$, $\underline{\mathbf{R}}$.
    \end{algorithmic}
\end{algorithm}

\textit{Remark:} Different from the decentralized algorithms in~\cite{zhaoCommunicationEfficientDecentralizedLinear2023}, which are mathematically equivalent to their centralized counterparts, the decentralized Algorithm~\ref{alg:dec_opt_overall} is not exactly equivalent to its centralized version, Algorithm~\ref{alg:opt_overall}. The inequivalence arises from the matrix inversion-free beamforming optimization introduced in Section~\ref{subsec:dec_w}, as well as the approximation of the term $\hat{\mathbf{W}}$ in~\eqref{eq:dec_opt_tx_position_w_norm}. Nevertheless, as will be demonstrated in Section~\ref{sec:sim}, the centralized and decentralized algorithms achieve nearly identical performance.

\subsection{Complexity Analysis}
For the centralized implementation given in Algorithm~\ref{alg:opt_overall}, the time complexity of a single BCA iteration is dominated by the updates of $\underline{\mathbf{W}}$ and $\mathbf{T}$, and is given by $\mathcal{O}(M^3 T_{\rm bis} + M^2 L_k^\mathtt{Tx} T_{\rm MM}^{\mathtt{Tx}})$. Here, $T_{\rm bis}$ denotes the number of iterations in the bisection search for updating $\underline{\mathbf{W}}$ described in Section~\ref{subsec:bisec_beamforming}, and $T_{\rm MM}^{\mathtt{Tx}}$ represents the average number of MM iterations required for updating $\mathbf{T}$. The complexity of the centralized implementation is proportional to $M^3$ and grows rapidly as the number of transmit FAs $M$ increases. 

In comparison, the complexity of the decentralized implementation at the CU is given by $\mathcal{O}(K^2 d^3 + N^3 K + T_{\rm MM}^{\mathtt{Rx}}(N^2 L_k^\mathtt{Rx} + N(L_k^\mathtt{Rx})^2))$, and the complexity at each DU is given by $\mathcal{O}(N K^2 d + K^2 d^2 + T_{\rm MM}^{\mathtt{Tx}}(M_c d L_k^\mathtt{Tx} + d(L_k^\mathtt{Tx})^2))$, where $T_{\rm MM}^{\mathtt{Tx}}$ and $T_{\rm MM}^{\mathtt{Rx}}$ represent the MM iterations for the update step of $\mathbf{T}$ and $\underline{\mathbf{R}}$, respectively. Notably, the complexities of the CU and each DU grow as a function of $M_c$, instead of $M$, and are significantly lower than those of the centralized implementation.

\section{Simulation}\label{sec:sim}
In this section, we evaluate the performance of FA-assisted MU-MIMO networks optimized using the proposed centralized BCA-based algorithm in Algorithm~\ref{alg:opt_overall} and its decentralized implementation in Algorithm~\ref{alg:dec_opt_overall}. The centralized and decentralized algorithms are represented as ``C'' and ``D'', respectively.

We denote the system with joint beamforming and transmit and receive FA position optimization as transmit and receive FA (TRFA). We compare the performance of TRFA with several baselines, specified as follows.
\begin{enumerate}
    \item \textbf{FPA}: The antenna arrays at the BS and users are fixed in position with a spacing of $\lambda / 2$.
    \item \textbf{Random-position antenna (RPA)}: The antennas at the BS and the users are FAs with random positions.
    \item \textbf{Transmit FA (TFA)}: The antennas at the users are fixed with a spacing of $\lambda / 2$. The antennas at the BS are FAs.
    \item \textbf{Receive FA (RFA)}: The antennas at the BS are fixed with a spacing of $\lambda / 2$, while the antennas at the users are FAs.
\end{enumerate}

\begin{table}[tb]
    \centering
    \caption{Key Simulation Parameters~\cite{zhangDiscerningEnhancingWeighted2023,dongMovableAntennaWireless2024,zhaoCommunicationEfficientDecentralizedLinear2023}}
    \begin{small}
    \begin{tabular}{|l|l|}
        \hline
        \textbf{Parameter} & \textbf{Value} \\ \hline
        Number of channel realizations & $S = 200$ \\
        Number of transmit FAs & $M = 64$ \\
        Number users &  $K = 6$ \\
        User priority & $\alpha_k = 1$ \\
        Number of FAs at each user &  $N = 4$     \\
        Number of parallel data streams &  $d = 4$      \\
        Number of DUs & $C = 4$ \\
        Carrier freqency & $f_c = 28$ GHz \\ 
        Carrier wavelength & $\lambda = 10.7$ mm \\
        Minimal distance between FAs & $D = \lambda / 2$ \\
        Noise power & $\sigma_k^2 = -90$ dBm \\
        Transmit power budget & $P_{\max} = 30$ dBm \\
        Miminum user distance from the BS & $d_{\min} = 100$ m \\
        Maximum user distance from the BS & $d_{\max} = 300$ m \\
        Distance from the BS to user $k$ & $d_k^2 \sim \mathcal{U}[d_{\min}^2, d_{\max}^2]$ \\
        Pathloss exponent & $\varrho = 3.67$ \\
        Pathloss at reference distance $d_0 = 1$ m & $T_0 = -61.4$ dB \\
        Elevation/Azimuth AoD & $\theta_{kq}^\mathtt{Tx}, \phi_{kq}^\mathtt{Tx} \sim \mathcal{U}[0, \pi)$ \\
        Elevation/Azimuth AoA & $\theta_{kq}^\mathtt{Rx}, \phi_{kq}^\mathtt{Rx} \sim \mathcal{U}[0, \pi)$ \\
        Number of transmit/receive paths & $L_k^\mathtt{Tx} = L_k^\mathtt{Rx} = 3$ \\
        \hline
    \end{tabular}
    \end{small}%
    \label{tab:sim_param}
\end{table}

Unless otherwise specified, the key simulation parameters follow the settings in Table~\ref{tab:sim_param}. The pathloss of user $k$ is calculated as $\kappa(d_k) = T_0 (d_k / d_0)^{-\varrho}$, and the PRM is diagonal with entries following $[\mathbf{\Sigma}_k]_{qq} \sim \mathcal{CN}(0, \kappa(d_k)/L)$. Without the box-constrained movement mode, all transmit (and receive) FAs can move within a shared cuboid region, where the edge length is $\rho\lambda\sqrt{M}$ (and $\rho\lambda\sqrt{N}$), and the height is $2\rho\lambda$. With the box-constrained movement mode, the specific movable regions of each FA are detailed in Section~\ref{subsec:bca_box}.

\subsection{Impact of Box-Constrained Movement Mode}\label{subsec:sim_box}
\begin{figure}[tb]
    \centering
    \includegraphics[width=0.45\textwidth]{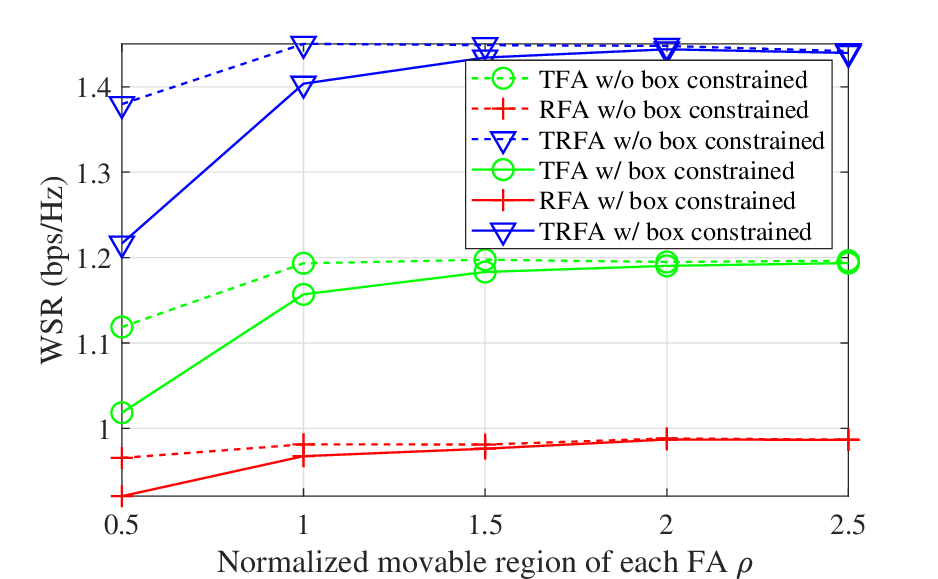}
    \caption{WSR comparison with and without the box-constrained movement mode.}
    \label{fig:bound}
\end{figure}

We evaluate the impact of the box-constrained movement mode on system performance. The parameter $\rho$ is used to control the size of the movable region for each FA. Specifically, when $\rho = 0.5$, each FA is restricted to movement along the $y$-axis for the box-constrained movement mode. As $\rho$ increases, the movable region expands, allowing greater flexibility. Fig.~\ref{fig:bound} presents the simulation results with $M = 16$. The WSR performance gap between systems with and without the box-constrained movement mode decreases as $\rho$ increases. Notably, when $\rho \geq 2$, the performance gap remains below $1\%$. This indicates that once the movable region of each FA is sufficiently large, the impact of the box-constrained movement mode on performance is negligible. This is because a larger $\rho$ provides sufficient spatial DoF under both movement modes, resulting in comparable capabilities to avoid deep fading channels. Based on this observation, we only test the performance with the box-constrained movement mode, and set $\rho = 2$ for all subsequent simulations.

\subsection{Convergence Behavior}
\begin{figure}[tb]
    \centering
    \includegraphics[width=0.45\textwidth]{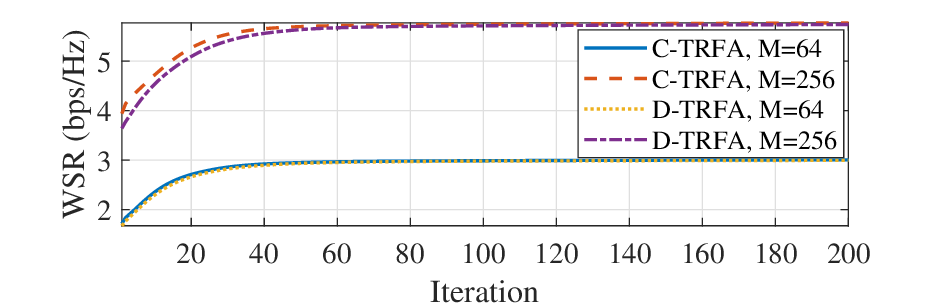}
    \caption{Convergence behaviors of the proposed algorithms.}
    \vspace{-0.3cm}
    \label{fig:conv}
\end{figure}

Fig.~\ref{fig:conv} illustrates the convergence behaviors of our proposed algorithms. Regardless of the number of transmit FAs $M$, all algorithms converge within $80$ iterations. It is also observed that the decentralized implementation converges slightly slower than the centralized approach, especially in the initial iterations. Nonetheless, the performance gap becomes negligible after convergence. 

\subsection{Performance Comparison}
\subsubsection{Comparison with Prior Art~\cite{fengWeightedSumRateMaximization2024}} 
\begin{figure}[tb]
    \centering
    \includegraphics[width=0.4\textwidth]{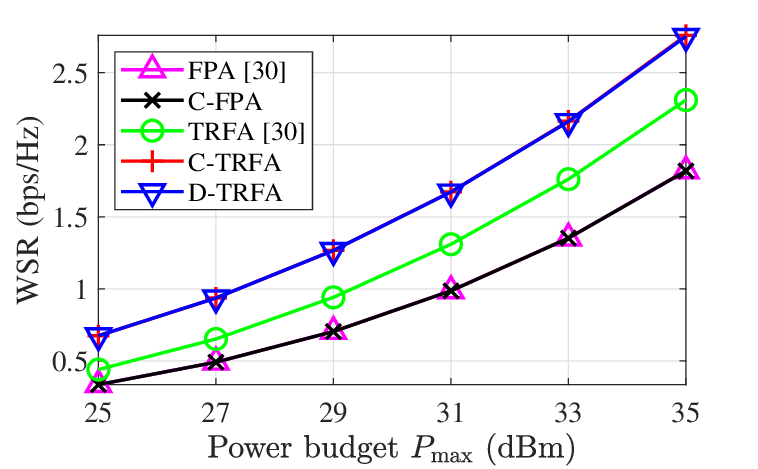}
    \caption{WSR comparison with~\cite{fengWeightedSumRateMaximization2024} under MU-MISO.}
    \vspace{-0.5cm}
    \label{fig:miso}
\end{figure}
We compare the WSR performance of the proposed centralized and decentralized algorithms with the conventional method in~\cite{fengWeightedSumRateMaximization2024}. Since the method in~\cite{fengWeightedSumRateMaximization2024} is designed for MU-MISO systems, we simplify our system by setting $K = 1$ and $d = 1$. The simulation results are presented in Fig.~\ref{fig:miso}. For FPA, the WMMSE algorithm in~\cite{fengWeightedSumRateMaximization2024} and the FP-based algorithm proposed in this paper achieve nearly identical performance, which is expected due to the equivalence between WMMSE and FP~\cite{shen2018fractional,shen2018fractional2}. For TRFA, both the proposed centralized and decentralized algorithms outperform the conventional method. The WSR gain is attributed to the proposed parallel MM algorithm, which constructs tighter surrogate functions and enables more effective optimization.

\subsubsection{Comparison with Baselines}
\begin{table*}[tb]
    \centering
    \caption{WSR Performance Comparison}
    \begin{small}
        \begin{tabular}{|c|c|cccccccccc|}
            \hline
            \multirow{2}{*}{$M$} &
              \multirow{2}{*}{\begin{tabular}[c]{@{}c@{}}$P_{\max}$\\ (dBm)\end{tabular}} &
              \multicolumn{10}{c|}{WSR (bps/Hz)} \\ \cline{3-12} 
             &
               &
              \multicolumn{1}{c|}{C-FPA} &
              \multicolumn{1}{c|}{D-FPA} &
              \multicolumn{1}{c|}{C-RPA} &
              \multicolumn{1}{c|}{D-RPA} &
              \multicolumn{1}{c|}{C-TFA} &
              \multicolumn{1}{c|}{D-TFA} &
              \multicolumn{1}{c|}{C-RFA} &
              \multicolumn{1}{c|}{D-RFA} &
              \multicolumn{1}{c|}{C-TRFA} &
              D-TRFA \\ \hline
            \multirow{2}{*}{$16$} &
              $30$ &
              \multicolumn{1}{c|}{$0.682$} &
              \multicolumn{1}{c|}{$0.682$} &
              \multicolumn{1}{c|}{$0.640$} &
              \multicolumn{1}{c|}{$0.641$} &
              \multicolumn{1}{c|}{$1.10$} &
              \multicolumn{1}{c|}{$1.10$} &
              \multicolumn{1}{c|}{$0.908$} &
              \multicolumn{1}{c|}{$0.909$} &
              \multicolumn{1}{c|}{$1.33$} &
              $1.34$ \\ \cline{2-12} 
             &
              $40$ &
              \multicolumn{1}{c|}{$3.02$} &
              \multicolumn{1}{c|}{$3.03$} &
              \multicolumn{1}{c|}{$2.98$} &
              \multicolumn{1}{c|}{$2.98$} &
              \multicolumn{1}{c|}{$4.01$} &
              \multicolumn{1}{c|}{$3.99$} &
              \multicolumn{1}{c|}{$3.64$} &
              \multicolumn{1}{c|}{$3.64$} &
              \multicolumn{1}{c|}{$4.52$} &
              $4.51$ \\ \hline
            \multirow{2}{*}{$64$} &
              $30$ &
              \multicolumn{1}{c|}{$1.76$} &
              \multicolumn{1}{c|}{$1.76$} &
              \multicolumn{1}{c|}{$1.69$} &
              \multicolumn{1}{c|}{$1.69$} &
              \multicolumn{1}{c|}{$2.51$} &
              \multicolumn{1}{c|}{$2.51$} &
              \multicolumn{1}{c|}{$2.04$} &
              \multicolumn{1}{c|}{$2.05$} &
              \multicolumn{1}{c|}{$2.87$} &
              $2.87$ \\ \cline{2-12} 
             &
              $40$ &
              \multicolumn{1}{c|}{$6.53$} &
              \multicolumn{1}{c|}{$6.53$} &
              \multicolumn{1}{c|}{$6.58$} &
              \multicolumn{1}{c|}{$6.58$} &
              \multicolumn{1}{c|}{$7.83$} &
              \multicolumn{1}{c|}{$7.74$} &
              \multicolumn{1}{c|}{$7.15$} &
              \multicolumn{1}{c|}{$7.15$} &
              \multicolumn{1}{c|}{$8.47$} &
              $8.38$ \\ \hline
            \multirow{2}{*}{$256$} &
              $30$ &
              \multicolumn{1}{c|}{$4.06$} &
              \multicolumn{1}{c|}{$4.06$} &
              \multicolumn{1}{c|}{$4.00$} &
              \multicolumn{1}{c|}{$4.00$} &
              \multicolumn{1}{c|}{$5.10$} &
              \multicolumn{1}{c|}{$5.07$} &
              \multicolumn{1}{c|}{$4.40$} &
              \multicolumn{1}{c|}{$4.41$} &
              \multicolumn{1}{c|}{$5.58$} &
              $5.55$ \\ \cline{2-12} 
             &
              $40$ &
              \multicolumn{1}{c|}{$13.2$} &
              \multicolumn{1}{c|}{$13.2$} &
              \multicolumn{1}{c|}{$13.6$} &
              \multicolumn{1}{c|}{$13.6$} &
              \multicolumn{1}{c|}{$14.4$} &
              \multicolumn{1}{c|}{$14.2$} &
              \multicolumn{1}{c|}{$14.0$} &
              \multicolumn{1}{c|}{$14.0$} &
              \multicolumn{1}{c|}{$15.3$} &
              $15.0$ \\ \hline
        \end{tabular}
    \end{small}%
    \label{tab:perf}
\end{table*}

We compare the WSR performance of TRFA, optimized by both the proposed centralized and decentralized implementations, against several baselines under various configurations, as summarized in Table~\ref{tab:perf}. The performance of C-TRFA and D-TRFA is consistently the highest across all configurations, demonstrating that the proposed BCA-based algorithm can exploit the spatial DoF of the system. The effectiveness of the proposed MM algorithm is validated by comparing the WSR values of TRFA and RPA. The WSR of RPA is similar to that of FPA, indicating that random FA position adjustments hardly provide any performance gain. In contrast, TRFA achieves significantly higher WSR than FPA, demonstrating that the proposed MM algorithm effectively optimizes FA positions to enhance performance. Compared with the centralized implementation, the decentralized implementation achieves a similar performance. For FPA, the WSR of the decentralized algorithm is no worse than that of the centralized algorithm. For TFA, the maximum WSR loss of the decentralized implementation is $1.41\%$ when $M = 256$ and $P_{\max} = 40$ dBm. For TRFA, the maximum WSR loss of the decentralized implementation is $2.00\%$ when $M = 256$ and $P_{\max} = 40$ dBm as well.

\subsubsection{Impact of Power Budget and Number of Users}
\begin{figure}[tb]
    \centering
    \includegraphics[width=0.4\textwidth]{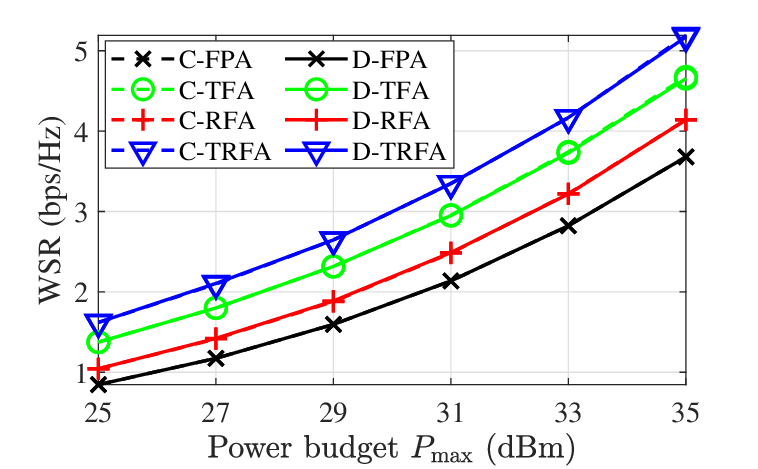}
    \caption{WSR versus the power budget $P_{\max}$.}
    \vspace{-0.5cm}
    \label{fig:power}
\end{figure}
\begin{figure}[tb]
    \centering
    \includegraphics[width=0.4\textwidth]{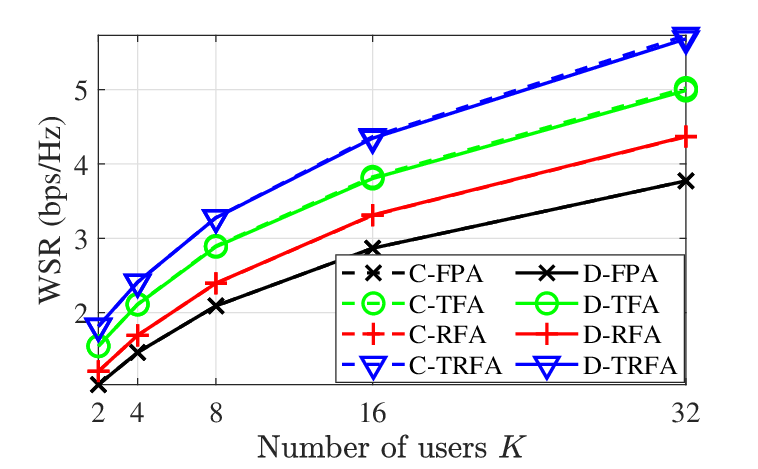}
    \caption{WSR versus the number of users $K$.}
    \vspace{-0.5cm}
    \label{fig:user}
\end{figure}
Fig.~\ref{fig:power} illustrates the WSR performance with different transmit power budgets. The WSR of all systems increases significantly with a higher transmit power budget, and TRFA consistently outperforms the baselines. The WSR of the decentralized implementation is similar to that of their centralized counterparts. The improved WSR performance of TRFA is attributed to the ability of FAs to dynamically reconstruct the channel, thereby enhancing the receive SINR under a fixed transmit power budget. As a result, FAs can significantly reduce the required transmit power to achieve a target performance. By fixing the WSR at $2$ bps/Hz, the transmit power budget can be reduced by around $4$ dB, demonstrating the superior performance of the FA-assisted MU-MIMO system.

The impact of the number of users is shown in Fig.~\ref{fig:user}. The WSR performance of TRFA consistently exceeds that of all baselines, and the decentralized implementation achieves a similar WSR as the centralized implementation. More importantly, as the number of users $K$ increases, the performance gap between TRFA and the baseline methods becomes more pronounced. In conventional MU-MIMO systems with FPAs, a larger number of users leads to more severe MUI, thereby degrading system performance. In contrast, the dynamically repositioning of FAs allows effective MUI mitigation, which in turn enhances the overall system capacity.

\subsection{Robust Analysis}
\begin{figure}[tb]
    \centering
    \includegraphics[width=0.4\textwidth]{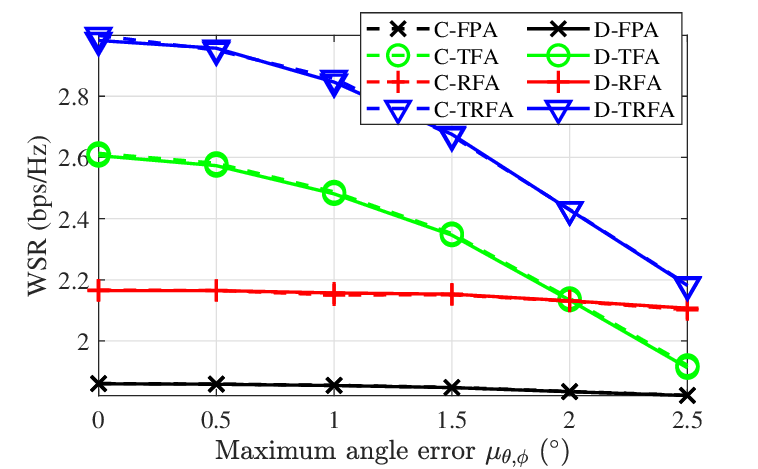}
    \caption{WSR versus the AoA/AoD error $\mu_{\theta, \phi}$.}
    \label{fig:robust_ang}
\end{figure}

Throughout the paper, we assume that CSI is perfectly known at the BS. This assumption, however, may not hold in practice due to channel estimation errors. In this part, we evaluate the effect of CSI errors on the system performance. First, we evaluate the impact of AoA/AoD errors on WSR. Denote the estimated elevation and azimuth AoD as $\tilde{\theta}_{kq}^{\mathtt{Tx}}$ and $\tilde{\phi}_{kq}^{\mathtt{Tx}}$, respectively, and the estimated elevation and azimuth AoA as $\tilde{\theta}_{kq}^{\mathtt{Rx}}$ and $\tilde{\phi}_{kq}^{\mathtt{Rx}}$, respectively. The difference between the estimated AoA/AoD and the ground truth AoA/AoD follows a uniform distribution, i.e., $\tilde{\theta}_{kq}^{\mathtt{Tx}} - \theta_{kq}^{\mathtt{Tx}} \sim \mathcal{U}\left[-\mu_{\theta, \phi}, \mu_{\theta, \phi}\right]$, $\tilde{\phi}_{kq}^{\mathtt{Tx}} - \phi_{kq}^{\mathtt{Tx}} \sim \mathcal{U}\left[-\mu_{\theta, \phi}, \mu_{\theta, \phi}\right]$, $\tilde{\theta}_{kq}^{\mathtt{Rx}} - \theta_{kq}^{\mathtt{Rx}} \sim \mathcal{U}\left[-\mu_{\theta, \phi}, \mu_{\theta, \phi}\right]$, and $\tilde{\phi}_{kq}^{\mathtt{Rx}} - \phi_{kq}^{\mathtt{Rx}} \sim \mathcal{U}\left[-\mu_{\theta, \phi}, \mu_{\theta, \phi}\right]$, where $\mu_{\theta, \phi}$ is the maximum AoA/AoD error. The simulation result, shown in Fig.~\ref{fig:robust_ang}, demonstrates that TRFA and TFA are more sensitive to the AoA/AoD errors than FPA and RPA. The reason is that inaccurate AoA/AoD may mislead the FAs to position themselves on undesirable channels, leading to performance degradation.

\begin{figure}[tb]
    \centering
    \includegraphics[width=0.4\textwidth]{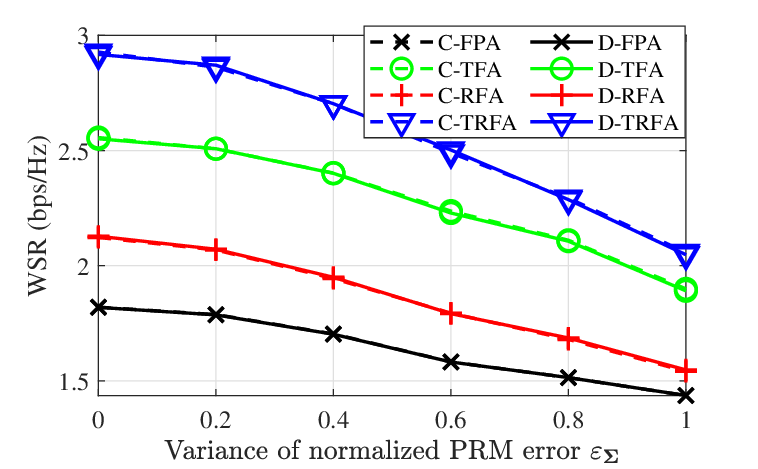}
    \caption{WSR versus the PRM error $\varepsilon_{\mathbf{\Sigma}}$.}
    \label{fig:robust_prm}
\end{figure}

Then, we evaluate the impact of PRM errors on WSR. We represent the estimated PRM as $\tilde{\mathbf{\Sigma}}_k$. The normalized difference of each entry between the estimated PRM and the ground truth PRM follows a CSCG distribution, i.e., $\frac{[\tilde{\mathbf{\Sigma}}_k]_{qq} - [\mathbf{\Sigma}_k]_{qq}}{\lvert[\mathbf{\Sigma}_k]_{qq}\rvert} \sim \mathcal{CN}\left(0, \varepsilon_{\mathbf{\Sigma}}\right)$, where $\varepsilon_{\mathbf{\Sigma}}$ is the PRM error. As shown in Fig.~\ref{fig:robust_prm}, all schemes are similarly robust to the PRM error. Even with very high PRM errors, i.e., $\varepsilon_{\mathbf{\Sigma}} = 1$, the WSR of TRFA is still the highest among all schemes. The phenomenon consolidates the robustness of the proposed algorithm.

\textit{Remark:} The proposed algorithm exhibits higher sensitivity to AoA/AoD errors compared to PRM errors. This is because small-scale fading, which FAS primarily exploits to enhance WSR, is more significantly influenced by AoA/AoD parameters. The simulation results further indicate that the proposed algorithm has a higher demand on the accuracy of AoA/AoD estimation than on PRM estimation.

\subsection{Computational Efficiency of Decentralized Implementation}
\begin{table}[tb]
    \caption{CPU Time Saved Compared with Centralized \\ Implementation (\%)}%\vspace{0.2cm}
    \centering
    \begin{small}
        \begin{tabular}{|c|c|c|c|c|c|}
            \hline
            $M$                    & $C$  & FPA      & TFA      & RFA      & TRFA     \\ \hline
            \multirow{2}{*}{$64$}  & $4$  & $95.5\%$ & $67.3\%$ & $77.8\%$ & $65.6\%$ \\ \cline{2-6} 
                                    & $16$ & $96.1\%$ & $77.1\%$ & $78.4\%$ & $73.1\%$ \\ \hline
            \multirow{2}{*}{$256$} & $4$  & $98.8\%$ & $87.6\%$ & $95.0\%$ & $87.0\%$ \\ \cline{2-6} 
                                    & $16$ & $99.1\%$ & $89.4\%$ & $95.2\%$ & $88.6\%$ \\ \hline
        \end{tabular}
    \end{small}%
    \vspace{-0.2cm}
    \label{tab:time_dr}
\end{table}
We quantify the computational efficiency of the proposed algorithm by measuring central processing unit (CPU) time. Specifically, the total CPU time is recorded for the centralized algorithm, while for the decentralized algorithm under the DBP architecture, the CPU time is computed as the sum of the CU's CPU time and the maximum running time among all DUs. As shown in Table~\ref{tab:time_dr}, the DBP architecture significantly reduces computation time by at least $65.6\%$ compared with the centralized algorithm across all system configurations. This improvement stems from the parallel processing capability of the DBP architecture, where each DU independently solves a smaller-scale problem. For a fixed number of transmit FAs $M$, increasing the number of DUs $C$ enhances parallelism, thereby saving more computational time.
\section{Conclusion}\label{sec:conclusion}
In this paper, we investigated the joint beamforming and antenna position optimization problem for WSR maximization in FA-assisted MU-MIMO networks. To tackle the inherent coupling between beamforming matrices and antenna positions, we employed matrix FP techniques to decouple the problem and adopted the BCA framework to solve the resulting subproblems. For antenna position optimization, we proposed a novel parallel MM algorithm that enables simultaneous updates of all FA positions. To further reduce computational overhead, we developed a decentralized implementation based on the DBP architecture. Simulation results demonstrate that the proposed parallel MM algorithm significantly outperforms existing FA position optimization methods in terms of WSR performance. Moreover, FA-assisted MU-MIMO networks optimized by our algorithms achieve significant WSR gains across various setups compared with conventional MU-MIMO systems. The decentralized implementation achieves substantial reductions in computation time while maintaining performance that is nearly identical to its centralized counterpart.
Additionally, we analyzed the robustness of the proposed algorithm to different types of channel uncertainty. The results reveal that the algorithm is more sensitive to AoA/AoD errors than to PRM errors, underscoring the importance of accurate angle estimation in FA-assisted MU-MIMO networks.

\appendix
\subsection{Derivation of $\nabla_{\vect\left(\mathbf{T}\right)} f_{\rm Quad}\left(\mathbf{T}\right)$}\label{appendix:dfdt}
The entries of $\nabla_{\vect\left(\mathbf{T}\right)} f_{\rm Quad}\left(\mathbf{T}\right)$ are computed using the matrix chain rule:
\begin{align}\label{eq:opt_position_chain}
    \frac{\partial f_{\rm Quad}}{\partial p_m^\mathtt{Tx}} 
    &= \sum_{k=1}^{K} \trace\left[ \left( \frac{\partial f_{\rm Quad}}{\partial \mathbf{G}_k} \right)^\transpose \frac{\partial \mathbf{G}_k}{\partial p_m^\mathtt{Tx}} 
    + \frac{\partial \mathbf{G}_k^\hermconj}{\partial p_m^\mathtt{Tx}} \left( \frac{\partial f_{\rm Quad}}{\partial \mathbf{G}_k^\hermconj} \right)^\transpose \right] \nonumber \\
    &= 2 \sum_{k=1}^{K} \Re\left\{ \trace\left( \mathbf{D}_k^\mathtt{Tx} \frac{\partial \mathbf{G}_k}{\partial p_m^\mathtt{Tx}} \right) \right\},
\end{align}
where $p \in \{x, y, z\}$ and $\mathbf{D}_k^\mathtt{Tx}$ is already given in~\eqref{eq:opt_tx_position_D}. Therefore, to compute $\frac{\partial f_{\rm Quad}}{\partial p_m^\mathtt{Tx}}$, it suffices to derive $\frac{\partial \mathbf{G}_k}{\partial p_m^\mathtt{Tx}}$:
\begin{equation}\label{eq:opt_position_dGF_dxyz}
    \frac{\partial \mathbf{G}_k}{\partial p_m^\mathtt{Tx}} 
    = \Big[\underbrace{\mathbf{0}, \dots, \mathbf{0}}_{m-1}, \frac{\partial \mathbf{g}_k\left(\mathbf{t}_m\right)}{\partial p_m^\mathtt{Tx}}, \underbrace{\mathbf{0}, \dots, \mathbf{0}}_{M-m}\Big] \in \mathbb{C}^{L_k^\mathtt{Tx} \times M}.
\end{equation}
The $q$-th element of the partial derivative $\frac{\partial \mathbf{g}_k(\mathbf{t}_m)}{\partial p_m^\mathtt{Tx}}$ can be compactly expressed as
\begin{align}\label{eq:opt_tx_position_dg_dxyz}
    \left[ \frac{\partial \mathbf{g}_k(\mathbf{t}_m)}{\partial \mathbf{t}_m^\transpose} \right]_q 
    &= \left[ \left[ \frac{\partial \mathbf{g}_k(\mathbf{t}_m)}{\partial x_m^{\mathtt{Tx}}} \right]_q, 
    \left[ \frac{\partial \mathbf{g}_k(\mathbf{t}_m)}{\partial y_m^{\mathtt{Tx}}} \right]_q, 
    \left[ \frac{\partial \mathbf{g}_k(\mathbf{t}_m)}{\partial z_m^{\mathtt{Tx}}} \right]_q \right] \nonumber \\
    &= \jmath \frac{2\pi}{\lambda} \left( \mathbf{g}_{kq}^\mathtt{Tx} \right)^\transpose 
    \exp\left( \jmath \frac{2\pi}{\lambda} \rho_{kq}^\mathtt{Tx}(\mathbf{t}_m) \right).
\end{align}
By substituting~\eqref{eq:opt_tx_position_dg_dxyz} into~\eqref{eq:opt_position_dGF_dxyz}, and then combining~\eqref{eq:opt_tx_position_D} and~\eqref{eq:opt_position_dGF_dxyz} into~\eqref{eq:opt_position_chain}, we obtain the final expression of $\nabla_{\vect\left(\mathbf{T}\right)} f_{\rm Quad}\left(\mathbf{T}\right)$ given in~\eqref{eq:opt_tx_position_dfdxyz}.

\subsection{Derivation of $\delta^{\mathtt{Tx}}$}\label{appendix:delta}
As indicated in~\eqref{eq:opt_tx_position_delta_bound}, the constant $\delta^{\mathtt{Tx}}$ is constructed such that its value the upper bound of the maximum eigenvalue of Hessian matrix $\nabla_{\vect\left(\mathbf{T}\right)}^2 f_{\rm Quad}(\mathbf{T})$. Since the calculation of eigenvalue is computationally expensive, we continue to find an upper bound of the maximum eigenvalue of the Hessian matrix to derive the closed-form expression for $\delta^{\mathtt{Tx}}$:
\begin{align}\label{eq:opt_tx_position_delta_ineq}
    &\lambda_{\max}\left(\nabla^2_{\vect\left(\mathbf{T}\right)}f_{\rm Quad}(\mathbf{T})\right) \nonumber \\
    \leq& \Big\lVert\nabla^2_{\vect\left(\mathbf{T}\right)}f_{\rm Quad}(\mathbf{T})\Big\rVert_{\infty} \nonumber \\ 
    =& \underset{\substack{1\leq m\leq M \\ p\in\{x, y, z\}}}{\max} \sum_{j=1}^M \left(\left\lvert\frac{\partial^2 f_{\rm Quad}}{\partial p_m^{\mathtt{Tx}} \partial x_j^{\mathtt{Tx}}}\right\rvert + \left\lvert\frac{\partial^2 f_{\rm Quad}}{\partial p_m^{\mathtt{Tx}} \partial y_j^{\mathtt{Tx}}}\right\rvert + \left\lvert\frac{\partial^2 f_{\rm Quad}}{\partial p_m^{\mathtt{Tx}} \partial z_j^{\mathtt{Tx}}}\right\rvert\right).
\end{align}%
First, we derive the expression of $\frac{\partial^2 f_{\rm Quad}}{\partial p_m^\mathtt{Tx} \partial p_j^{\prime,\mathtt{Tx}}}$:
\begin{align}\label{eq:opt_tx_position_d2f_dx2}
    \frac{\partial^2 f_{\rm Quad}}{\partial p_m^\mathtt{Tx} \partial p_j^{\prime, \mathtt{Tx}}} = 2\sum_{k=1}^K\Re&\left\{-[\hat{\mathbf{W}}]_{mj}\frac{\partial \mathbf{g}_k^\hermconj\left(\mathbf{t}_j\right)}{\partial p_j^{\prime, \mathtt{Tx}}} \hat{\mathbf{\Sigma}}_k^{\mathtt{Tx}} \frac{\partial \mathbf{g}_k\left(\mathbf{t}_m\right)}{\partial p_m^\mathtt{Tx}} \right.\nonumber \\
    &\phantom{-}\left.+ \delta_{mj} [\mathbf{D}_k^\mathtt{Tx}]_m \frac{\partial^2 \mathbf{g}_k\left(\mathbf{t}_m\right)}{\partial p_m^\mathtt{Tx} \partial p_j^{\prime, \mathtt{Tx}}}\right\},
\end{align}
where $p, p^\prime\in\{x, y, z\}$ and $\delta_{mj}$ denotes the Kronecker symbol. Since $\hat{\mathbf{W}}$ and $\hat{\mathbf{\Sigma}}_k^{\mathtt{Tx}}$ are constants w.r.t. $\mathbf{T}$, we then find the upper bounds of $\big\lvert\big[\frac{\partial \mathbf{g}_k(\mathbf{t}_m)}{\partial p_m^{\mathtt{Tx}}}\big]_q\big\rvert$, $\big\lvert\big[\frac{\partial^2 \mathbf{g}_k(\mathbf{t}_m)}{\partial p_m^{\mathtt{Tx}} \partial p_j^{\prime, \mathtt{Tx}}}\big]_q\big\rvert$, and $\lVert[\mathbf{D}_k^{\mathtt{Tx}}]\rVert_2$ for \emph{all} possible $\mathbf{T}$. From~\eqref{eq:opt_tx_position_dg_dxyz}, we note that
\begin{equation}\label{eq:opt_tx_position_dgdx_ub}
    \left\lvert\left[\frac{\partial \mathbf{g}_k(\mathbf{t}_m)}{\partial p_m^{\mathtt{Tx}}}\right]_q\right\rvert \leq \frac{2\pi}{\lambda} \quad \text{and} \quad \left\lvert\left[\frac{\partial^2 \mathbf{g}_k(\mathbf{t}_m)}{\partial p_m^{\mathtt{Tx}} \partial p_j^{\prime, \mathtt{Tx}}}\right]_q\right\rvert \leq \frac{4\pi^2}{\lambda^2}.
\end{equation}
The upper bound of $\big\lVert[\mathbf{D}_k^{\mathtt{Tx}}]_m\big\rVert_2$ can be calculated by triangle inequality:
\begin{align}\label{eq:opt_tx_position_D_ub}
    \Big\lVert[\mathbf{D}_k^\mathtt{Tx}]_m\Big\rVert_2 \leq& \sqrt{\alpha_k} \Big\lVert[\overline{\mathbf{W}}_k]_m\left(\mathbf{I}+\overline{\mathbf{\Gamma}}_k\right)\overline{\mathbf{\Phi}}_k^\hermconj\overline{\mathbf{F}}_k^\hermconj\mathbf{\Sigma}_k^\hermconj\Big\rVert_2 \nonumber \\
    &+ \sqrt{M L_k^\mathtt{Tx}} \Big\lVert[\hat{\mathbf{W}}]_m\Big\rVert_2 \Big\lVert\hat{\mathbf{\Sigma}}_k^{\mathtt{Tx}}\Big\rVert_2.
\end{align}
Combining~\eqref{eq:opt_tx_position_dgdx_ub} and~\eqref{eq:opt_tx_position_D_ub}, we use the triangle inequality to~\eqref{eq:opt_tx_position_d2f_dx2} to derive the upper bound of $\big\lvert\frac{\partial^2 f_{\rm Quad}}{\partial p_m^\mathtt{Tx} \partial p_j^{\prime,\mathtt{Tx}}}\big\rvert$:
\begin{small}
\begin{align}\label{eq:opt_tx_position_d2f_dx2_bound}
    \left\lvert\frac{\partial^2 f_{\rm Quad}}{\partial p_m^\mathtt{Tx} \partial p_j^{\prime,\mathtt{Tx}}}\right\rvert \leq& \frac{8\pi^2}{\lambda^2}\sum_{k=1}^K \left(\left(\Big\lvert[\hat{\mathbf{W}}]_{mj}\Big\rvert+\sqrt{M}\delta_{mj}\Big\lVert[\hat{\mathbf{W}}]_m\Big\rVert_2\right) \lVert\hat{\mathbf{\Sigma}}_k^{\mathtt{Tx}}\rVert_2\nonumber \right.\\
    &\left.+ \delta_{mj}\sqrt{\frac{\alpha_k}{L_k^\mathtt{Tx}}}\Big\lVert[\overline{\mathbf{W}}_k]_m\left(\mathbf{I}+\overline{\mathbf{\Gamma}}_k\right)\overline{\mathbf{\Phi}}_k^\hermconj\overline{\mathbf{F}}_k^\hermconj\mathbf{\Sigma}_k^\hermconj\Big\rVert_2\right)L_k^\mathtt{Tx}
\end{align}
\end{small}%
Finally, we plug the inequality~\eqref{eq:opt_tx_position_d2f_dx2_bound} into~\eqref{eq:opt_tx_position_delta_ineq}, and obtain a upper bound of $\lambda_{\max}\left(\nabla^2_{\vect\left(\mathbf{T}\right)}f_{\rm Quad}(\mathbf{T})\right)$ for \emph{all} possible $\mathbf{T}$. The result is assigned to $\delta^{\mathtt{Tx}}$ and shown in~\eqref{eq:opt_tx_position_delta}.

\subsection{Proof of Lemma~\ref{lemma:ub}}\label{appendix:upper_boundedness}
\begin{IEEEproof}
According to the properties of matrix Lagrangian dual transform~\cite[Theorem~4]{shenOptimizationMIMODevicetoDevice2019} and the matrix quadratic transform~\cite[Theorem~3]{shenOptimizationMIMODevicetoDevice2019}, the objective function $f_{\rm Quad}(\underline{\mathbf{W}}, \mathbf{T}, \underline{\mathbf{R}}, \underline{\mathbf{\Gamma}}, \underline{\mathbf{\Phi}})$ is upper bounded by the WSR $R$, as long as the variables are updated according to Algorithm~\ref{alg:opt_overall}:
\begin{equation}
    f_{\rm Quad}\left(\underline{\mathbf{W}}, \mathbf{T}, \underline{\mathbf{R}}, \underline{\mathbf{\Gamma}}, \underline{\mathbf{\Phi}}\right) \leq f_{\rm Lag}\left(\underline{\mathbf{W}}, \mathbf{T}, \underline{\mathbf{R}}, \underline{\mathbf{\Gamma}}\right) \leq R.
\end{equation}
To find the upper bound of $f_{\rm Quad}\left(\underline{\mathbf{W}}, \mathbf{T}, \underline{\mathbf{R}}, \underline{\mathbf{\Gamma}}, \underline{\mathbf{\Phi}}\right)$, we only need to find the upper bound of $R_k$.

To begin with, we derive the upper bounds of $\lVert\mathbf{W}_k\rVert_{\rm F}^2$, $\lVert\mathbf{H}_k(\mathbf{T}, \mathbf{R}_k)\rVert_{\rm F}^2$, and $\lVert\mathbf{M}_k^{-1}\rVert_{\rm F}$, which are given by
\begin{subequations}
    \begin{equation}
        \Big\lVert\mathbf{W}_k\Big\rVert_{\rm F}^2 \leq P_{\max},
    \end{equation}
    \begin{equation}
        \Big\lVert\mathbf{H}_k(\mathbf{T}, \mathbf{R}_k)\Big\rVert_{\rm F}^2 = \lVert\mathbf{F}_k^\hermconj(\mathbf{R}_k)\mathbf{\Sigma}_k\mathbf{G}_k(\mathbf{T})\rVert_{\rm F}^2 \leq MN(L_k^\mathtt{Tx} L_k^\mathtt{Rx})^2,
    \end{equation}
    and
    \begin{equation}
        \Big\lVert\mathbf{M}_k^{-1}\Big\rVert_{\rm F} = \left[\sum_{n=1}^N \lambda_n^{-2}(\mathbf{M}_k)\right]^{1/2} \leq \frac{\sqrt{N}}{\sigma_k^2},
    \end{equation}
\end{subequations}
respectively, where $\lambda_n(\mathbf{M}_k)$ is the $n$-th eigenvalue of $\mathbf{M}_k$. Based on the above results, we then derive the upper bound of $R_k$ as follows:
\begin{align}
    R_k &= \log\det\left(\mathbf{I} + \mathbf{W}_k^\hermconj\mathbf{H}_k^\hermconj\mathbf{M}_k^{-1}\mathbf{H}_k\mathbf{W}_k\right) \nonumber \\
    &= \sum_{i=1}^d \log\left(1 + \lambda_i(\mathbf{W}_k^\hermconj\mathbf{H}_k^\hermconj\mathbf{M}_k^{-1}\mathbf{H}_k\mathbf{W}_k)\right) \nonumber \\
    &\leq d\log\left(1 + \lambda_{\max}(\mathbf{W}_k^\hermconj\mathbf{H}_k^\hermconj\mathbf{M}_k^{-1}\mathbf{H}_k\mathbf{W}_k)\right) \nonumber \\
    &\leq d\log\left(1 + \lVert\mathbf{W}_k\rVert_{\rm F}^2 \lVert\mathbf{H}_k\rVert_{\rm F}^2 \lVert\mathbf{M}_k^{-1}\rVert_{\rm F}\right) \nonumber \\
    &= d\log\left(1 + MN^{3/2}(L_k^\mathtt{Tx} L_k^\mathtt{Rx})^2 P_{\max}/\sigma_k^2\right).
\end{align}
Therefore, for all $\underline{\mathbf{W}}, \mathbf{T}, \underline{\mathbf{R}}, \underline{\mathbf{\Gamma}}, \underline{\mathbf{\Phi}}$ satisfying~\eqref{opt:power_constraint}--\eqref{opt:rx_coupling_constraint}, the desired $R_{\max}$ can be constructed as
\begin{equation}
    R_{\max} = d\sum_{k=1}^K \alpha_k \log\left(1 + MN^{3/2}(L_k^\mathtt{Tx} L_k^\mathtt{Rx})^2 P_{\max}/\sigma_k^2\right).
\end{equation}
\end{IEEEproof}
\bibliographystyle{IEEEtran}

\bibliography{IEEEabrv, reference}

\begin{IEEEbiography}[{\includegraphics[width=1in,height=1.25in,clip,keepaspectratio]{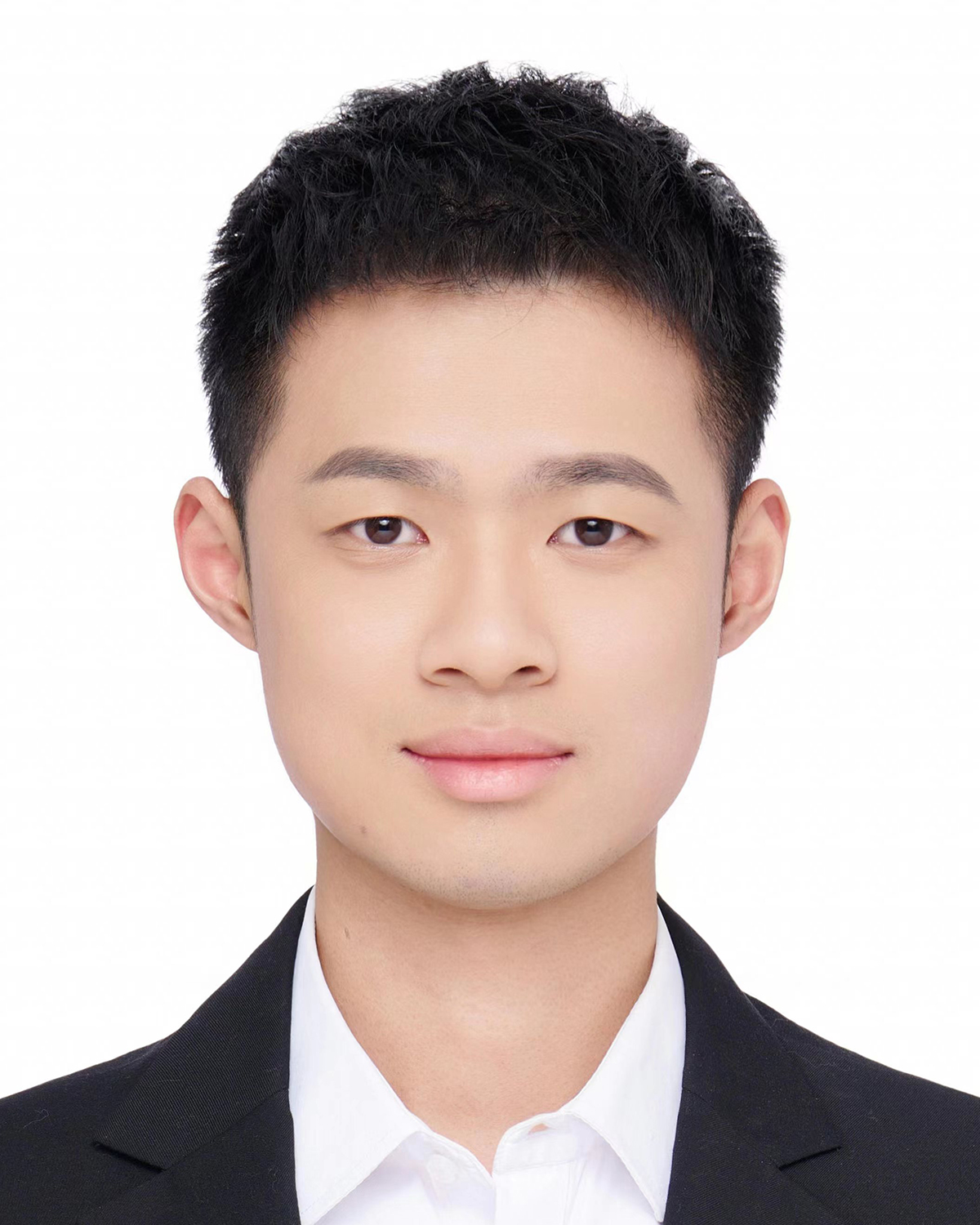}}]
    {Tianyi Liao} (Graduate Student Member, IEEE) received the B.Eng. degree in Information Engineering from Southeast University (SEU), Nanjing, China, in 2024. He is currently pursuing the Ph.D. degree in the Department of Electronic and Computer Engineering at the Hong Kong University of Science and Technology (HKUST) under the supervision of Prof. Khaled B. Letaief. His research interests include fluid-antenna systems (FASs), reconfigurable antennas, and mathematical optimization. He received the China National Scholarship and the Hong Kong Ph.D. Fellowship Scheme (HKPFS) in 2024.
\end{IEEEbiography}

\begin{IEEEbiography}[{\includegraphics[width=1in,height=1.25in,clip,keepaspectratio]{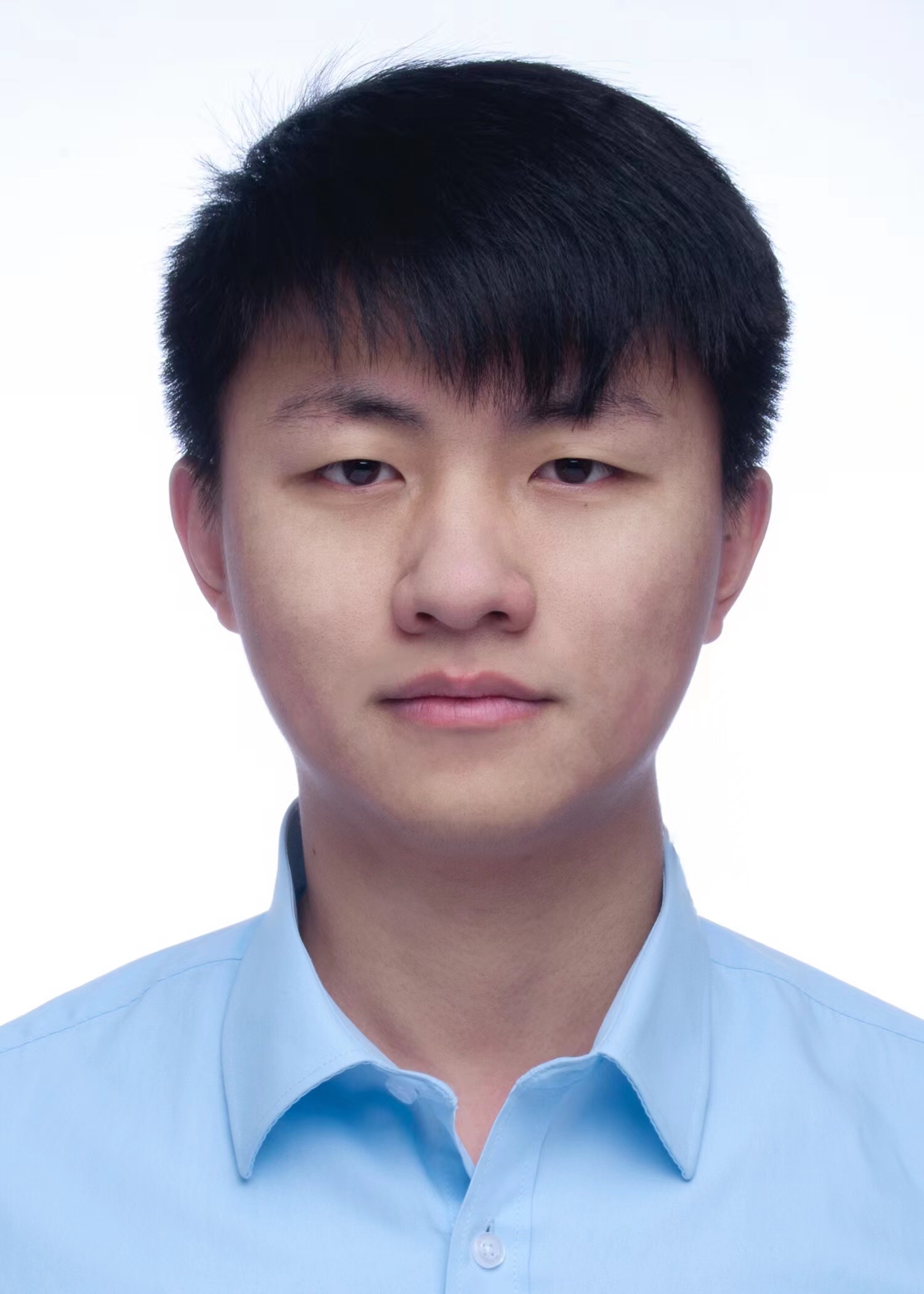}}]
    {Wei Guo} (Member, IEEE) received the B.Eng. degree in electrical engineering from the University of Electronic Science and Technology of China, Chengdu, China, in 2017, and the Ph.D. degree in computer and information engineering with The Chinese University of Hong Kong, Shenzhen, China, in 2023. He is currently working as Postdoctoral Fellow with the Department of Electronic and Computer Engineering at The Hong Kong University of Science and Technology. His current research interests include integrated communications and AI and edge AI. He was a recipient of an Exemplary Reviewers 2020 of \textsc{IEEE Wireless Communications Letters}.
\end{IEEEbiography}

\begin{IEEEbiography}[{\includegraphics[width=1in,height=1.25in,clip,keepaspectratio]{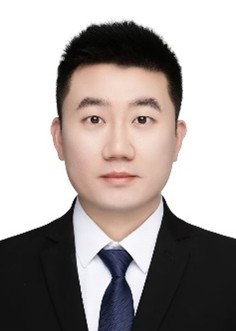}}]
    {Hengtao He} (Member, IEEE) received the B.S. degree in communications engineering from Nanjing University of Science and Technology, China, in 2015, and the Ph.D. degree in information and communications engineering from Southeast University, China, in 2021. From 2018 to 2020, he was a Visiting Student with the Department of Electrical and Computer Engineering at the Georgia Institute of Technology, Atlanta, GA, USA. From 2021 to 2025, he worked as a Post-Doctoral Fellow and Research Assistant Professor under the scheme of VPRDO with the Department of Electronic and Computer Engineering at The Hong Kong University of Science and Technology. He is currently a Professor at Southeast University.

    His areas of interests currently include machine learning for wireless communications, edge AI and wireless sensing. He was the recipient of the first prize of the Natural Science Award of the Chinese Institute of Electronics in 2023, Best Ph.D. Thesis Award of the Chinese Institute of Communications and Jiangsu Province in 2022, and Top 2\% Scientists Worldwide 2023--2025 by Stanford University. He is the Editor of \textsc{IEEE Wireless Communications Letters}.
\end{IEEEbiography}

\begin{IEEEbiography}[{\includegraphics[width=1in,height=1.25in,clip,keepaspectratio]{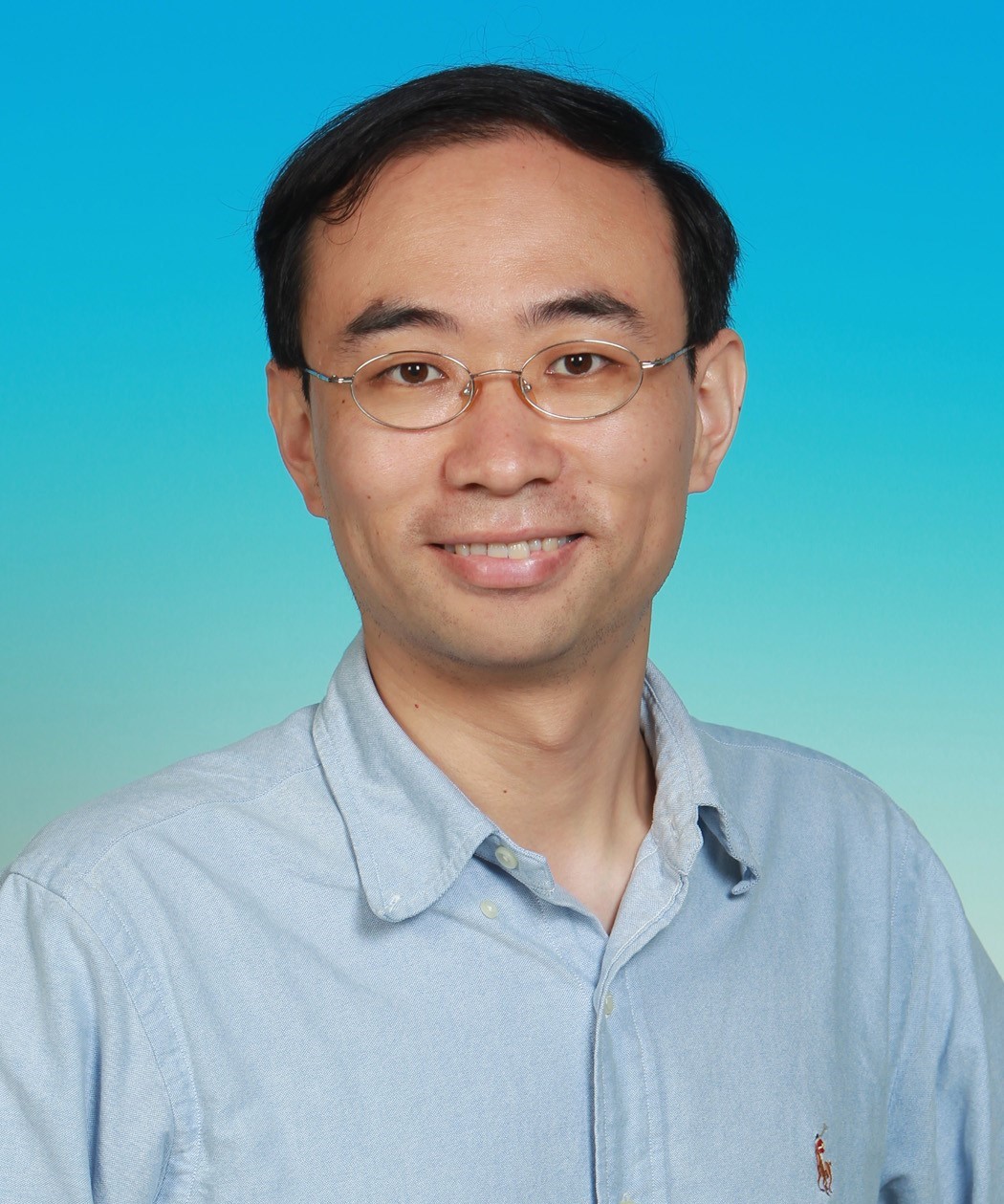}}]
    {Shenghui Song} (Senior Member, IEEE) Dr. S.H. Song is now an Associate Professor jointly appointed by the Division of Integrative Systems and Design (ISD) and the Department of Electronic and Computer Engineering (ECE) at the Hong Kong University of Science and Technology (HKUST). His research is primarily in the areas of Wireless Communications and Machine Learning with current focus on Information Theory, Integrated Sensing and Communication, Distributed Intelligence, Semantic Communications, and Machine Learning for Communications. Dr. Song is an editorial board member of \textit{Entropy}. He was named the Exemplary Reviewer for \textsc{IEEE Communications Letters}. He served as the Tutorial Program Co-Chairs of the 2022 IEEE International Mediterranean Conference on Communications and Networking, and the Technical Program Chairs of the International Conference on 6G Communications Networking and Signal Processing, 2023 and 2024.

    Dr. Song is also interested in the research on Engineering Education and served as an Associate Editor for the \textsc{IEEE Transactions on Education}. He has won several teaching awards at HKUST, including the Michael G. Gale Medal for Distinguished Teaching in 2018, the Best Ten Lecturers in 2013, 2015, and 2017, the School of Engineering Distinguished Teaching Award in 2012, the Teachers I Like Award in 2013, 2015, 2016, and 2017, and the MSc (Telecom) Teaching Excellent Appreciation Award for 2020--21 and 2022--23. Dr. Song was one of the honorees of the Third Faculty Recognition at HKUST in 2021.
\end{IEEEbiography}

\begin{IEEEbiography}[{\includegraphics[width=1in,height=1.25in,clip,keepaspectratio]{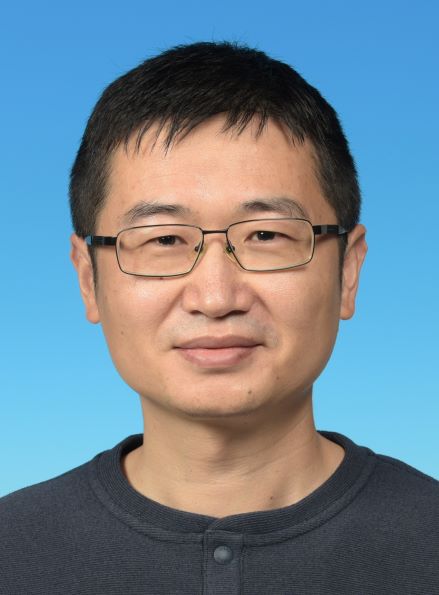}}]
    {Jun Zhang} (Fellow, IEEE) received the Ph.D. degree in Electrical and Computer Engineering from the University of Texas at Austin in 2009. He is a Professor in the Department of Electronic and Computer Engineering (ECE) and Associate Director of the Computer Engineering (CPEG) Program at the Hong Kong University of Science and Technology. His research interests include integrated communications and AI, generative AI, and edge AI systems. He is an IEEE Fellow and was an IEEE ComSoc Distinguished Lecturer (2023-2024).

    Dr. Zhang co-authored/co-edited five books including \textit{Fundamentals of LTE} (Prentice-Hall, 2010). He is a co-recipient of several best paper awards, including the 2025 IEEE Communications Society Katherine Johnson Young Author Best Paper Award, the 2021 Best Survey Paper Award of the IEEE Communications Society, the 2019 IEEE Communications Society \& Information Theory Society Joint Paper Award, and the 2016 Marconi Prize Paper Award in Wireless Communications. He also received the 2016 IEEE ComSoc Asia-Pacific Best Young Researcher Award. He is currently an Area Editor of \textsc{IEEE Transactions on Wireless Communications} (leading the area of Machine Learning and Artificial Intelligence) and \textsc{IEEE Transactions on Machine Learning in Communications and Networking} (leading the area of Distributed Learning and AI at the Network Edge). He served as a symposium co-chair for IEEE Wireless Communications and Networking Conference (WCNC) 2011 and 2026, IEEE International Conference on Communications (ICC) 2021, and a TPC co-chair for The IEEE Hong Kong 6G Wireless Summit 2023, 2024. 
\end{IEEEbiography}

\begin{IEEEbiography}[{\includegraphics[width=1in,height=1.25in,clip,keepaspectratio]{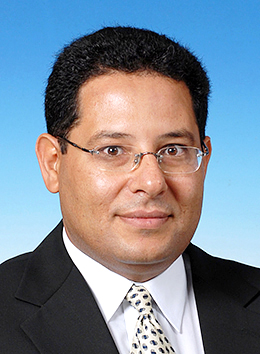}}]
    {Khaled B. Letaief} (Fellow, IEEE) is a globally recognized leader in wireless communications and networks, with a research focus that spans artificial intelligence, integrated sensing and communication, mobile cloud and edge computing, federated learning, and 6G systems. His prolific contributions include over 700 publications, which have garnered more than 62,700 citations with an h-index of 110. He holds 15 inventions, including 11 U.S. patents.

    Dr. Letaief is a distinguished member of several esteemed organizations, including the United States National Academy of Engineering, IEEE Fellow, and Fellow of the Hong Kong Institution of Engineers. He is also a member of the Hong Kong Academy of Engineering Sciences. His research excellence has earned him recognition as an ISI Highly Cited Researcher by Thomson Reuters, and he was named one of the top 30 Most Influential Scholars in AI and the Internet of Things in 2020.

    His accolades include numerous prestigious awards, such as the 2024 IEEE James Evans Avant Garde Award, 2024 Distinguished Purdue University Alumni Award, 2022 IEEE Edwin Howard Armstrong Achievement Award, and 2021 IEEE Communications Society Best Survey Paper Award. He has also received the 2019 Joint Paper Award from the IEEE Communications Society and Information Theory Society, the 2016 IEEE Marconi Prize Award in Wireless Communications, and over 20 IEEE Best Paper Awards.

    Since 1993, Dr. Letaief has been a faculty member at The Hong Kong University of Science and Technology (HKUST), where he has held multiple leadership roles, including Senior Advisor to the President, Acting Provost, Head of the Electronic and Computer Engineering Department, and Director of the Hong Kong Telecom Institute of Information Technology. He served as Chair Professor and Dean of Engineering at HKUST and, from 2015 to 2018, was Provost at Hamad Bin Khalifa University in Qatar, where he played a key role in establishing a research-intensive university in collaboration with renowned institutions like Northwestern University, Carnegie Mellon University, Cornell, and Texas A\&M.

    Dr. Letaief is celebrated for his dedicated service to professional societies and IEEE, having held numerous leadership positions, including Division Director and member of the IEEE Board of Directors, founding Editor-in-Chief of the esteemed \textsc{IEEE Transactions on Wireless Communications}, and President of the IEEE Communications Society from 2018 to 2019, the leading global organization for communications professionals.

    He earned his B.S. degree with distinction in Electrical Engineering from Purdue University in December 1984, followed by an M.S. and Ph.D. in Electrical Engineering from the same institution in August 1986 and May 1990, respectively. In 2022, he received an honorary Ph.D. from the University of Johannesburg, South Africa.
\end{IEEEbiography}
    
\end{document}